\documentclass{statsoc}
\usepackage{latexsym, epsfig, amssymb, amsmath, graphicx, mathrsfs}
\usepackage{times}
\usepackage{anysize}
\usepackage[OT1]{fontenc}

\marginsize{1.2in}{1in}{0.5in}{1.5in} %
\makeatletter
\def\singlespace{\def\baselinestretch{1}\@normalsize}

\newtheorem{condition}{Condition}
\newtheorem{lemma}{Lemma}
\newtheorem{theorem}{Theorem}
\newtheorem{definition}{Definition}
\renewcommand{\theequation}{
\arabic{equation}%
}

\makeatother

\newcommand{\bb}{\mbox{\bf b}}

\newcommand{\bv}{\mbox{\bf v}}

\newcommand{\bx}{\mbox{\bf x}}
\newcommand{\by}{\mbox{\bf y}}

\newcommand{\bX}{\mbox{\bf X}}

\newcommand{\bone}{\mbox{\bf 1}}
\newcommand{\bzero}{\mbox{\bf 0}}
\newcommand{\bveps}{\mbox{\boldmath $\varepsilon$}}

\newcommand{\bbeta}{\mbox{\boldmath $\beta$}}
\newcommand{\bdelta}{\mbox{\boldmath $\delta$}}

\newcommand{\bleta}{\mbox{\boldmath $\eta$}}

\newcommand{\bmu}{\mbox{\boldmath $\mu$}}

\newcommand{\hbbeta}{\widehat\bbeta}

\newcommand{\hbeta}{\widehat\beta}

\newcommand{\Sig}{\mathbf{\Sigma}}
\newcommand{\veps}{\varepsilon}
\newcommand{\tr}{\mathrm{tr}}
\newcommand{\diag}{\mathrm{diag}}

\newcommand{\sgn}{\mathrm{sgn}}
\newcommand{\supp}{\mathrm{supp}}

\def\t{^T}

\title[Thresholded Regression and Shrinkage Effect]
{High-Dimensional Thresholded Regression and Shrinkage Effect}
\author[Z. Zheng, Y. Fan and J. Lv]{Zemin Zheng, Yingying Fan and Jinchi Lv}
\address{University of Southern California, Los Angeles, USA}
\email{zeminzhe@usc.edu, fanyingy@marshall.usc.edu, jinchilv@marshall.usc.edu}

\begin{document}

\begin{abstract}
High-dimensional sparse modeling via regularization provides a powerful tool for analyzing large-scale data sets and obtaining meaningful, interpretable models. The use of nonconvex penalty functions shows advantage in selecting important features in high dimensions, but the global optimality of such methods still demands more understanding. In this paper, we consider sparse regression with hard-thresholding penalty, which we show to give rise to thresholded regression. This approach is motivated by its close connection with the $L_0$-regularization, which can be unrealistic to implement in practice but of appealing sampling properties, and its computational advantage. Under some mild regularity conditions allowing possibly exponentially growing dimensionality, we establish the oracle inequalities of the resulting regularized estimator, as the global minimizer, under various prediction and variable selection losses, as well as the oracle risk inequalities of the hard-thresholded estimator followed by a further $L_2$-regularization. The risk properties exhibit interesting shrinkage effects under both estimation and prediction losses. We identify the optimal choice of the ridge parameter, which is shown to have simultaneous advantages to both the $L_2$-loss and prediction loss. These new results and phenomena are evidenced by simulation and real data examples.

\end{abstract}

\section{Introduction} \label{Sec1}
The advances of information technologies in the past few decades have made it much easier than before to collect large amount of data over a wide spectrum of dimensions in different fields. As a powerful tool of sparse modeling and variable selection, regularization methods have been widely used to analyze large-scale data sets and produce meaningful, interpretable models. Depending on the type of penalty functions used, the regularization methods can be grouped as two classes: convex ones and nonconvex ones. A typical example of convex penalty is the $L_1$-penalty which gives rise to the $L_1$-regularization methods such as the Lasso (Tibshirani, 1996) and Dantzig selector (Candes and Tao, 2007). The convexity of these methods makes the implementation efficient and facilitates the theoretical analysis. In a seminal paper, Bickel, Ritov and Tsybakov (2009) established the oracle inequalities of both the Lasso and Dantzig selector under various prediction and estimation losses and, in particular, proved their asymptotic equivalence under certain conditions.

Despite their convexity and popularity, it has become a well-known phenomenon that convex regularization methods can suffer from the bias issue that is inherited from the convexity of the penalty function. This issue can deteriorate the power of identifying important covariates and the efficiency of estimating their effects in high dimensions. To attenuate this issue, Fan and Li (2001) initiated the general framework of nonconcave penalized likelihood with nonconvex penalties including the proposed smoothly clipped absolute deviation (SCAD) penalty, and showed that the oracle properties can hold for a wide class of nonconvex regularization methods. Other nonconvex regularization methods include the bridge regression using the $L_q$-penalty for $0 < q < 1$ (Frank and Friedman, 1993), the minimax concave penalty (MCP) (Zhang, 2010), and the smooth integration of counting and absolute deviation (SICA) penalty (Lv and Fan, 2009). A main message of these works is that nonconvex regularization can be beneficial in selecting important covariates in high dimensions.

Although there is a growing literature on nonconvex regularization methods, some important questions still remain. As an important step, most of existing studies for these methods focus on some local minimizer with appealing properties due to their general nonconvexity. 
Yet the 
properties of 
the global minimizer need more delicate analysis, 
and the global theory may depend on the specific form of regularization. 
A natural question is whether 
the oracle inequalities hold for nonconvex regularization methods as for the $L_1$-regularization methods (Candes and Tao, 2007; Bickel, Ritov and Tsybakov, 2009), 
and the logarithmic factor of dimensionality that appears commonly in the oracle inequalities is optimal. In the 
problem of Gaussian mean estimation, there is a well-known phenomenon of Stein's shrinkage effect (Stein, 1956; James and Stein, 1961) stating that the maximum likelihood estimator or least-squares estimator may no longer be optimal in risk under the quadratic loss in multiple dimensions. Thus another natural question is whether similar shrinkage effects hold for these 
methods under both estimation and prediction losses.

In this paper, we intend to provide some partial answers to the aforementioned questions, with a focus on one particular member of the nonconvex family, the hard-thresholding penalty. The $L_0$-regularization, which amounts to the best subset regression, motivated different forms of regularization. This method was proved to enjoy the oracle risk inequalities under the prediction loss in Barron, Birge and Massart (1999). It is, however, unrealistic to implement in practice due to its combinatorial computational complexity. As an alternative to the $L_0$-penalty, the hard-thresholding penalty is continuous with a fixed, finite maximum concavity which controls the computational difficulty. We show that both approaches 
give rise to a thresholded regression. As is well known in the wavelets literature, the hard-thresholding regularization is equivalent to the $L_0$-regularization in the case of orthonormal design matrix. This connection motivates us to fully investigate the approach of hard-thresholding regularization.

The main contributions of this paper are twofold. First, we establish comprehensive global properties of the hard-thresholding regularization, including the oracle inequalities 
under various prediction and variable selection losses 
and the nonoptimality of the logarithmic factor of dimensionality. 
Second, we show that the hard-thresholding regularization followed by a further $L_2$-regularization 
enjoys interesting Stein's shrinkage effects in terms of risks under both estimation and prediction losses. The identified optimal choice of the ridge parameter is revealed to have simultaneous advantages to both the $L_2$-loss and prediction loss, which result builds an interesting connection between model selection and prediction. These new results and phenomena provide further insights into the hard-thresholding regularization method.

The rest of the paper is organized as follows. Section \ref{Sec2} presents the thresholded regression with the hard-thresholding penalty and $L_0$-penalty, and their hard-thresholding property. We establish the global properties of thresholded regression under various prediction and variable selection losses and unveil Stein's shrinkage effects for both estimation and prediction losses, with optimal choices of the ridge parameter, in Section \ref{Sec3}. Section \ref{Sec5} discusses briefly the implementation of thresholded regression. We provide several simulation and real data examples in Section \ref{Sec6}. All technical details are relegated to the Appendix and Supplementary Material.

\section{Thresholded regression} \label{Sec2}
To address the questions raised in the Introduction, we focus our attention on the 
linear regression model
\begin{equation} \label{001}
\by = \bX \bbeta + \bveps,
\end{equation}
where $\by = (y_1, \cdots, y_n)\t$ is an $n$-dimensional response vector, $\bX = (\bx_1, \cdots, \bx_p)$ is an $n \times p$ deterministic design matrix consisting of $p$ covariate vectors, 
$\bbeta = (\beta_1, \cdots, \beta_p)\t$ is a $p$-dimensional regression coefficient vector, and $\bveps = (\veps_1, \cdots, \veps_n)\t$ is an $n$-dimensional error vector. 
The goal of variable selection is to consistently recover the true underlying sparse model $\supp(\bbeta_0) = \{j: \beta_{0, j} \neq 0, 1\leq j \leq p\}$ for the true regression coefficient vector $\bbeta_0 =  (\beta_{0,1}, \cdots, \beta_{0,p})\t$ in model (\ref{001}), and to estimate the $s = \|\bbeta_0\|_0$ nonzero regression coefficients $\beta_{0, j}$'s.

To produce a sparse estimate of $\bbeta_0$, we consider the approach of penalized least squares which minimizes
\begin{equation} \label{e001}
Q(\bbeta) = (2n)^{-1} \|\by - \bX \bbeta\|_2^2 + \|p_\lambda(\bbeta)\|_1,
\end{equation}
the penalized residual sum of squares with penalty function $p_\lambda(t)$. Here we use the compact notation $p_\lambda(\bbeta) = p_\lambda(|\bbeta|) = (p_\lambda(|\beta_1|), \cdots, p_\lambda(|\beta_p|))\t$ with $|\bbeta| = (|\beta_1|, \cdots, |\beta_p|)\t$. The penalty function $p_\lambda(t)$, defined on $t \in [0, \infty)$ and indexed by $\lambda \geq 0$, is assumed to be increasing in both $t$ and $\lambda$ with $p_\lambda(0) = 0$, indicating that the amount of regularization increases with the magnitude of the parameter and the regularization parameter $\lambda$. To align all covariates to a common scale, we assume that each covariate vector $\bx_j$ is rescaled to have $L_2$-norm $n^{1/2}$, matching that of the constant covariate $\bone$ for the intercept. 
See, for example, the references mentioned in the Introduction for the specific forms of various penalty functions that have been proposed 
for sparse modeling.

As elucidated in the Introduction, we focus on the hard-thresholding penalty
\begin{equation} \label{003}
p_{H, \lambda}(t) = \frac{1}{2} \left[\lambda^2 - (\lambda - t)_+^2\right], \quad t \geq 0,
\end{equation}
which is closely related to the $L_0$-penalty $p_{H_0, \lambda}(t) = 2^{-1} \lambda^2 1_{\{t \neq 0\}}$, $t \geq 0$. It is well known that in the wavelets setting with the design matrix $\bX$ multiplied by $n^{-1/2}$ being orthonormal, that is, $n^{-1} \bX\t \bX = I_p$, the penalized least squares in (\ref{e001}) reduces to a componentwise minimization problem with $Q(\bbeta) = 2^{-1} \|\hbbeta_{\text{ols}} - \bbeta\|_2^2 + \|p_\lambda(\bbeta)\|_1$, where $\hbbeta_{\text{ols}} = n^{-1} \bX\t \by$ is the ordinary least-squares estimator. 
In this setting, the use of hard-thresholding penalty $p_{H, \lambda}(t)$ gives the componentwise hard-thresholding, which is of the form $z 1_{\{|z| > \lambda\}}$, on the ordinary least-squares estimator (Antoniadis, 1996). In contrast, the use of the $L_1$-penalty $p_\lambda(t) = \lambda t$ yields the soft-thresholding which is of the form $\sgn(z) (|z| - \lambda)_+$. When the $L_0$-penalty $p_{H_0, \lambda}(t)$ is used, an identical hard-thresholding rule to that by hard-thresholding penalty $p_{H, \lambda}(t)$ is obtained. We see that in the case of orthonormal design matrix, both approaches of hard-thresholding regularization and $L_0$-regularization are equivalent. This simple connection suggests that they may have more general connection, which motivates our study.

Moreover, the hard-thresholding penalty in (\ref{003}) is continuous and has fixed, finite maximum concavity
\begin{equation} \label{009}
\kappa(p_{H, \lambda}) = \sup_{0 < t_1 < t_2 <\infty} -\frac{p_{H, \lambda}'(t_2) - p_{H, \lambda}'(t_1)}{t_2 - t_1} = 1,
\end{equation}
which is related to the computational difficulty of the regularization method and gives rise to its computational advantage.
The computationally attractive method of hard-thresholding regularization indeed shares some similarity with $L_0$-regularization in the general case, as shown in the following lemma on the hard-thresholding property.

\begin{lemma}
\label{L1}
For both hard-thresholding penalty $p_{H, \lambda}(t)$ and $L_0$-penalty $p_{H_0, \lambda}(t)$, minimizing $Q(\bbeta)$ in (\ref{e001}) along the $j$-th coordinate with $1 \leq j \leq p$, at any $p$-vector $\bbeta_j$ with $j$-th component zero, gives the univariate global minimizer for that coordinate of the same form
$\hbeta(z) = z 1_{\{|z| > \lambda\}}$,
with $z = n^{-1} (\by - \bX \bbeta_j)\t  \bx_j$. 
\end{lemma}

The simple observation in Lemma \ref{L1} facilitates our technical analysis and enables us to derive parallel results for both methods. Since the global minimizer 
is necessarily the global minimizer along each coordinate, the characterization of each coordinate in the above lemma shows that the regularized estimators given by both methods 
are natural generalizations of the univariate hard-thresholded estimator. In this sense, we refer to both methods as thresholded regression using hard-thresholding. There is, however, no guarantee that both estimators are identical when $p > 1$. We show in Theorem \ref{Thm4} that the two methods 
can have similar oracle inequalities 
under various prediction and variable selection losses, 
which justifies a further connection between them.

\section{Global properties and shrinkage effects of thresholded regression} \label{Sec3}

\subsection{Technical conditions} \label{Sec3.1}
It is well known that high collinearity is commonly associated with large-scale data sets. High collinearity can lead to unstable estimation or even loss of model identifiability in regression problems. More specifically, there may exist another $p$-vector $\bbeta_1$ that is different from $\bbeta_0$ such that $\bX \bbeta_1$ is (nearly) identical to $\bX \bbeta_0$, when the dimensionality $p$ is large compared with the sample size $n$. Thus to ensure model identifiability and reduce model instability, it is necessary to control the size of sparse models, since it is clear from the geometric point of view that the collinearity among the covariates increases with the dimensionality. This idea was exploited in Donoho and Elad (2003) for the problem of sparse recovery, that is, the noiseless case of (\ref{001}). To ensure the identifiability of $\bbeta_0$, they introduced the concept of spark, denoted as $\text{spark}(\bX)$, for a design matrix $\bX$, which is defined as the smallest number $\tau$ such that there exists a linearly dependent subgroup of $\tau$ columns from $\bX$. In particular, $\bbeta_0$ is uniquely defined as long as $s < \text{spark}(\bX)/2$, which provides a basic condition for model identifiability. 

Since we are interested in variable selection in the presence of noise, we extend their concept of spark to the robust case as follows.

\begin{definition}
\label{Def1}
The robust spark $M = \emph{rspark}_c(\bX)$ of an $n \times p$ design matrix $\bX$ with bound $c$ is defined as the smallest 
number $\tau$ such that there exists a subgroup of $\tau$ columns from 
$n^{-1/2} \bX$ such that the corresponding submatrix has a singular value less than the given positive constant $c$.
\end{definition}

An equivalent representation of the robust spark $M = \text{rspark}_c(\bX)$ in Definition \ref{Def1} is the largest number $\tau$ such that the following inequality holds
\begin{equation} \label{010}
\min_{\|\bdelta\|_0 < \tau, \ \|\bdelta\|_2 = 1} n^{-1/2} \|\bX \bdelta\|_2 \geq c.
\end{equation}
This inequality provides a natural constraint on the collinearity for sparse models. In view of (\ref{010}), our robust spark condition of $s < M/2$, to be introduced in Condition \ref{cond2}, is in a similar spirit to the restricted eigenvalue condition in Bickel, Ritov and Tsybakov (2009). The restricted eigenvalue condition assumes (\ref{010}) with the $L_0$-norm constraint $\|\bdelta\|_0 < \tau$ replaced by the $L_1$-norm constraint of $\|\bdelta_{J_0^c}\|_1 \leq c_0 \|\bdelta_{J_0}\|_1$ for some positive constant $c_0$, where $J_0 \subset \{1, \cdots, p\}$ with $|J_0| \leq s'$, $J_0^c$ is the complement of $J_0$, and $\bdelta_A$ denotes a subvector of $\bdelta$ consisting of components with indices in a given set $A$. The robust spark condition of $s < M/2$ requires that (\ref{010}) holds for $\tau = 2 s + 1$. Since such an $L_0$-norm constraint generally defines a smaller subset than the above $L_1$-norm constraint 
for $s' = 2 s$, the robust spark condition can be weaker than the restricted eigenvalue condition.
It is easy to show that the robust spark $\mbox{rspark}_c(\bX)$ increases as $c$ decreases, and approaches the spark $\mbox{spark}(\bX)$ as $c \rightarrow 0+$. Thus $M$ can generally be any positive integer no larger than $n + 1$.

To ensure model identifiability and reduce the instability in estimated model, we consider the regularized estimator $\hbbeta$ on the union
of coordinate subspaces $\mathbb{S}_{M/2} = \{\bbeta \in \mathbb{R}^p: \|\bbeta\|_0 < M/2\}$, as exploited in Fan and Lv (2011) to characterize the global optimality of nonconcave penalized likelihood estimators. Thus throughout the paper, the regularized estimator $\hbbeta$ is defined as the global minimizer
\begin{equation} \label{002}
\hbbeta = \arg\min_{\bbeta \in \mathbb{S}_{M/2}} Q(\bbeta),
\end{equation}
where $Q(\bbeta)$ is defined in (\ref{e001}). When the size of sparse models exceeds $M/2$, that is, $\bbeta$ falls outside the space $\mathbb{S}_{M/2}$, there is generally no guarantee for model identifiability.

To facilitate our technical analysis, we make the following three regularity conditions.

\begin{condition} \label{cond1}
$\bveps \sim N(\bzero, \sigma^2 I_n)$ for some positive $\sigma$.
\end{condition}

\begin{condition} \label{cond2}
It holds that $s < M/2$, $s = o(n)$, and $b_0 = \min_{j \in \supp(\bbeta_0)} |\beta_{0, j}| > (\sqrt{16/c^2} \vee 1) c^{-1}c_2$ $\sqrt{(2s + 1) (\log \widetilde{p})/n}$, where 
$M$ is the robust spark of $\bX$ with bound $c$ given in Definition \ref{Def1}, $c_2 \geq \sqrt{10}\sigma$ is some positive constant, and $\widetilde{p} = n \vee p$.
\end{condition}

\begin{condition} \label{cond4}
$\|\bbeta_0\|_2$ is bounded from below by some positive constant and
$\max_{\|\bdelta\|_0 < M/2, \|\bdelta\|_2 = 1} n^{-1/2} \|\bX \bdelta\|_2$ $\leq c_3$ for some positive constant $c_3$.
\end{condition}

Condition \ref{cond1} is standard in the linear regression model. The Gaussian error distribution 
is assumed to simplify the technical arguments. The theoretical results continue to hold for other 
error distributions with possibly different probability bound in Theorem \ref{Thm4}; see, for example, Fan and Lv (2011) for more technical details. In particular, some numerical results for the $t$ error distribution are presented in Section \ref{Sec6.1.2}. The heavy-tailedness of the error distribution typically leads to lower probability for the prediction and variable selection bounds to hold.

The first part $s < M/2$ of Condition \ref{cond2} puts a sparsity constraint on the true model size $s$ that involves the robust spark given in Definition \ref{Def1}. As explained above, such a robust spark condition is needed to ensure model identifiability. We typically assume a diverging ratio of the sample size $n$ to the number of true covariates $s$, that is, $s = o(n)$, to obtain consistent estimation of $\bbeta_0$. The third part of Condition \ref{cond2} gives a lower bound on the minimum signal strength for model selection consistency.

Condition \ref{cond4}, which is only needed in Theorem \ref{Thm5}, facilitates the derivation of the oracle risk properties of the regularized estimator, which are stronger than the oracle inequalities in Theorem \ref{Thm4}. In particular, the first part of Condition \ref{cond4} assumes that the $L_2$-norm of 
$\bbeta_0$ is bounded from below, which is mild and sensible. The second part of Condition \ref{cond4} is a restricted-eigenvalue-type assumption and requires that the maximum singular value of each submatrix of $n^{-1/2} \bX$ by taking out less than $M/2$ columns is bounded from above.

\subsection{Global properties and shrinkage effects} \label{Sec3.2}
In view of Lemma \ref{L1}, the regularization parameter $\lambda$ determines the threshold level for both hard-thresholding penalty $p_{H, \lambda}(t)$ and $L_0$-penalty $p_{H_0, \lambda}(t)$. So a natural idea for ensuring the model selection consistency is choosing an appropriately large regularization parameter $\lambda$ to suppress 
all noise covariates and retain important ones. This approach is shown to be effective in the following theorem on the model selection consistency and oracle inequalities of thresholded regression.

\begin{theorem}
\label{Thm4}
Assume that Conditions \ref{cond1}--\ref{cond2} hold and $c^{-1} c_2 \sqrt{(2s + 1) (\log \widetilde{p})/n} < \lambda < b_0 (1 \wedge \sqrt{c^2/2})$. Then for both hard-thresholding penalty $p_{H, \lambda}(t)$ and $L_0$-penalty $p_{H_0, \lambda}(t)$, the regularized estimator $\hbbeta$ in (\ref{002}) satisfies that with probability at least $1 - (2/\pi)^{1/2} c_2^{-1} \sigma (\log \widetilde{p})^{-1/2} \widetilde{p}^{1 - c_2^2/(2 \sigma^2)} - (2/\pi)^{1/2} c_2'^{-1} \sigma s (\log n)^{-1/2}$ $n^{-c_2'^2/(2 \sigma^2)}$ for some positive constant $c_2' \geq \sqrt{2} \sigma$, it holds simultaneously that:
\begin{itemize}
\item[\emph{(a)}] \emph{(Model selection consistency)}. $\supp(\hbbeta) = \supp(\bbeta_0)$;

\item[\emph{(b)}] \emph{(Prediction loss)}. $n^{-1/2} ||\bX(\hbbeta-\bbeta_0)||_2 \leq  2c'_2c^{-1} \sqrt{s (\log n)/n}$;

\item[\emph{(c)}] \emph{(Estimation losses)}. $\|\hbbeta-\bbeta_0\|_q \leq 2c^{-2}c_2' s^{1/q} \sqrt{(\log n)/n}$ for $q \in [1, 2]$ and $\|\hbbeta-\bbeta_0\|_{\infty}$ is bounded by the same upper bound as for $q = 2$.
\end{itemize}
\end{theorem}

With the above choice of the regularization parameter $\lambda$, 
the prediction loss of the regularized estimator is within a logarithmic factor $(\log n)^{1/2}$ of that of the oracle estimator, which is referred to as the least-squares estimator on the true underlying sparse model. Theorem \ref{Thm4} also establishes the oracle inequalities of the regularized estimator under the $L_q$-estimation losses with $q \in [1, 2] \cup \{\infty\}$. These results hold simultaneously with significant probability that converges to one polynomially with sample size $n$, since $\widetilde{p} = n \vee p$. The dimensionality $p$ is allowed to grow up to exponentially fast with the sample size $n$, in view of the range for $\lambda$.

The key to deriving these rates is establishing the model selection consistency of the hard-thresholded estimator, that is, the exact recovery of the true underlying sparse model. Such a property enables us to construct a key event with significant probability, on which we can conduct delicate analysis. The suitable range of the regularization parameter is critical in this theorem since the lower bound on $\lambda$ is needed for suppressing all noise covariates and the upper bound on $\lambda$ is needed for retaining all true covariates, although this range is unknown to us in practice.

Theorem \ref{Thm4} builds on Lemma \ref{L1}, both of which share a common feature that the technical arguments apply equally to both hard-thresholding penalty and $L_0$-penalty. Thus, under conditions of Theorem \ref{Thm4}, the regularized estimators given by both hard-thresholding penalty $p_{H, \lambda}(t)$ and $L_0$-penalty $p_{H_0, \lambda}(t)$ are approximately asymptotically equivalent, that is, having the same convergence rates in the oracle inequalities under various prediction and variable selection losses. 
This formally justifies the motivation and advantage of studying the hard-thresholding regularization. In fact, their approximate asymptotic equivalence 
extends to the oracle risk inequalities under different prediction and variable selection losses. These results complement those on the oracle risk inequalities under the prediction loss in Barron, Birge and Massart (1999). Since it enjoys the same appealing properties as the $L_0$-regularization, the hard-thresholding regularization provides an attractive alternative to the $L_0$-regularization thanks to its computational advantage, as discussed in Section \ref{Sec2}.

As mentioned in the Introduction, many studies have contributed to the oracle inequalities for the $L_1$-regularization methods. For instance, Candes and Tao (2007) proved that the Dantzig selector can achieve a loss within a logarithmic factor of the dimensionality compared to that for the oracle estimator. Bunea, Tsybakov and Wegkamp (2007) established sparsity oracle inequalities for the Lasso estimator. Bickel, Ritov and Tsybakov (2009) derived parallel oracle inequalities for the Lasso estimator and Dantzig selector under the prediction loss and $L_q$-estimation losses with $q \in [1, 2]$. A common feature of these results is the appearance of some power of the logarithmic factor $\log p$ of the dimensionality $p$. In contrast, such a factor is replaced by the logarithmic factor $\log n$ of the sample size $n$ in our setting. This suggests the general nonoptimality of the logarithmic factor of dimensionality in the oracle inequalities when $p$ grows nonpolynomially with $n$. Our results are also related to other work on nonconvex regularization methods. Antoniadis and Fan (2001) obtained comprehensive oracle inequalities and universal thresholding parameters for a wide class of general penalty functions, in the wavelets setting. Zhang (2010) proved that the MCP estimator can attain certain minimax convergence rates 
for the estimation of regression coefficients in $L_q$-balls.

Although providing bounds on different estimation and prediction losses on an event with large probability, the oracle inequalities of the thresholded regression presented in Theorem \ref{Thm4} do not take into account its performance over the full sample space. Thus it is of interest to investigate a stronger property of oracle risk inequalities for thresholded regression, where the risk under a loss is its expectation over all realizations.
As shown in the proof of Theorem \ref{Thm4}, the hard-thresholded estimator $\hbbeta$ in (\ref{002}) on its support $\supp(\hbbeta)$ is exactly the ordinary least-squares estimator constructed using covariates in $\supp(\hbbeta)$. Motivated by such a representation, we consider a refitted estimator constructed by applying a further $L_2$-regularization to the thresholded regression
\begin{equation} \label{013}
\hbbeta_{\text{refitted}} = (\bX_1\t \bX_1 + \lambda_1 I_{s_1})^{-1} \bX_1\t \by,
\end{equation}
where $\bX_1$ is a submatrix of the design matrix $\bX$ consisting of columns in $\supp(\hbbeta)$, $s_1 = \|\hbbeta\|_0$, and $\lambda_1 \geq 0$ is 
the ridge parameter. In the special case of $\lambda_1 = 0$, the above refitted estimator $\hbbeta_{\text{refitted}}$ becomes the original hard-thresholded estimator $\hbbeta$ in (\ref{002}).

Let $\bX_0$ be a submatrix of the design matrix $\bX$ consisting of columns in $\supp(\bbeta_0)$ and $\bX_0\t \bX_0 = P\t D P$ an eigendecomposition with $P$ an orthogonal matrix and $D = \diag\{d_1, \cdots, d_s\}$. We show that Stein's shrinkage effects (Stein, 1956; James and Stein, 1961) also hold for the thresholded regression followed by the $L_2$-regularization in terms of risks under both estimation and prediction losses. These results are presented in the following theorem on the oracle risk inequalities of the $L_2$-regularized thresholded regression.

\begin{theorem}
\label{Thm5}
Assume that conditions of Theorem \ref{Thm4} and Condition \ref{cond4} hold. Then the $L_2$-regularized refitted estimator $\hbbeta_{\emph{refitted}}$ in (\ref{013}) satisfies  that:
\begin{itemize}
\item[\emph{(a)}] \emph{($L_2$-risk)}. The minimum $L_2$-risk $E\|\hbbeta_{\emph{refitted}}-\bbeta_0\|_2^2$ is attained at the optimal ridge parameter $\lambda_1 = \lambda_{1, \emph{opt}} = O(s \|\bbeta_0\|_2^{-2}) + O(s^2 n^{-1} \|\bbeta_0\|_2^{-4})$, with the leading term $O(s \|\bbeta_0\|_2^{-2})$ sandwiched between $s \sigma^2 \|\bbeta_0\|_2^{-2} (\lambda_{\min}/\lambda_{\max})^2$ and $s \sigma^2 \|\bbeta_0\|_2^{-2} (\lambda_{\max}/\lambda_{\min})^2$, and equals $O(s/n) + O(s^2 n^{-2} \|\bbeta_0\|_2^{-2})$ with the leading term $O(s/n)$ being $\sum_{j=1}^s (\lambda_{1, \emph{opt}}^2 b_j^2 + d_j \sigma^2)/(d_j+\lambda_{1, \emph{opt}})^2$;

\item[\emph{(b)}] \emph{($L_q$-risk)}. The minimum $L_q$-risk $E\|\hbbeta_{\emph{refitted}}-\bbeta_0\|_q^q$ equals $O(s/n^{q/2}) + O(s^2 \|\bbeta_0\|_2^{-2}/n^{q/2 + 1})$ for $q \in [1,2]$, and the minimum $L_\infty$-risk $E\|\hbbeta_{\emph{refitted}}-\bbeta_0\|_{\infty}$ equals $O(s^{1/2}/n^{1/2}) + O(s^{3/2} \|\bbeta_0\|_2^{-2}/n^{3/2})$;

\item[\emph{(c)}] \emph{(Prediction risk)}. The minimum prediction risk $n^{-1}E\|\bX(\hbbeta_{\emph{refitted}}-\bbeta_0)\|_2^2$ is attained at the optimal ridge parameter $\lambda_1 = \lambda_{1, \emph{opt}}' = O(s \|\bbeta_0\|_2^{-2}) + O(s^2 n^{-1} \|\bbeta_0\|_2^{-4})$, with the leading term $O(s \|\bbeta_0\|_2^{-2})$ sandwiched between $s \sigma^2 \|\bbeta_0\|_2^{-2} \lambda_{\min}/\lambda_{\max}$ and $s \sigma^2 \|\bbeta_0\|_2^{-2} \lambda_{\max}/\lambda_{\min}$, and equals $O(s/n) + O(s^2 n^{-2} \|\bbeta_0\|_2^{-2})$ with the leading term $O(s/n)$ being $n^{-1} \sum_{j=1}^s [(\lambda_{1, \emph{opt}}')^2 b_j^2 d_j + d_j^2 \sigma^2]/(d_j + \lambda_{1, \emph{opt}}')^2$,
\end{itemize}
where $(b_1,\cdots,b_s)\t = P \bbeta_{0, 1}$ with $\bbeta_{0, 1}$ a subvector of $\bbeta_0$ consisting of all nonzero components, and $\lambda_{\min}$ and $\lambda_{\max}$ are the smallest and largest eigenvalues of 
$\bX_0\t \bX_0$, respectively.
\end{theorem}

Although it has the well-known bias issue, the ridge regression applied after the thresholded regression is shown in Theorem \ref{Thm5} to be capable of improving both estimation and prediction, since the original hard-thresholded estimator is simply the refitted estimator with $\lambda_1 = 0$ and the minimum risks under the losses are attained at nonzero ridge parameters $\lambda_1$. Intuitively, the bias incurred by an appropriately small amount of $L_2$-regularization can be offset by the reduction in estimation variability, leading to improvement in the overall risks of the regularized estimator. This phenomenon can be clearly seen in the representative $L_2$-risk and prediction risk curves as a function of the ridge parameter $\lambda_1$ in Section \ref{Sec6}. The risks drop as $\lambda_1$ increases from zero, and start to rise after the minimum risks are attained.

The model selection consistency of the thresholded regression plays a key role in deriving the risk properties of the $L_2$-regularized refitted estimator. 
The optimal risks are attained at nontrivial ridge parameter $\lambda_1$ for both the $L_q$-loss and prediction loss. Since $s = o(n)$ by Condition \ref{cond2} and $\|\bbeta_0\|_2$ is bounded from below by some positive constant by Condition \ref{cond4}, we see that both optimal ridge parameters $\lambda_{1, \text{opt}}$ and $\lambda_{1, \text{opt}}'$ for the $L_2$-risk and prediction risk, respectively, are of the same leading order $O(s \|\bbeta_0\|_2^{-2})$. In particular, the leading term $O(s \|\bbeta_0\|_2^{-2})$ of the optimal ridge parameter for $L_2$-risk has a similar range to that of the optimal ridge parameter for prediction risk, differing by only a factor of $\lambda_{\max}/\lambda_{\min}$. Such a factor is the condition number of the Gram matrix $\bX_0\t \bX_0$ resulting from the true design matrix $\bX_0$. In view of (\ref{010}) and Condition \ref{cond4}, its condition number $\lambda_{\max}/\lambda_{\min}$ is sandwiched between $1$ and $c_3^2/c^2$.

It is interesting to observe that the optimal choices of the ridge parameter for both $L_2$-loss and prediction loss are of the same order $O(s \|\bbeta_0\|_2^{-2})$, 
which is proportional to the true model size $s$ and has an inverse relationship with $\|\bbeta_0\|_2$. This indicates that stronger signal leads to smaller optimal $L_2$-shrinkage. Thus the optimal ridge parameter has a simultaneous benefit on both the $L_2$-loss and prediction loss. Furthermore, the minimum $L_2$-risk and minimum prediction risk share the same order of $O(s/n)$, 
when the risks are minimized by the optimal ridge parameters. These risk properties demonstrate that Stein's shrinkage effects extend to the thresholded regression followed by the $L_2$-regularization under both estimation and prediction losses.

The idea of refitting has also been investigated in van de Geer, B\"{u}hlmann and Zhou (2011), who established bounds on the prediction loss and $L_q$-loss with $q \in [1, 2]$ for the thresholded Lasso estimator. 
The thresholded Lasso is a three-step procedure with the Lasso followed by hard-thresholding and an ordinary least-squares refitting, while the above $L_2$-regularized refitting is a two-step procedure with hard-thresholding and ordinary least-squares refitting automatic in thresholded regression. A main difference is that our study focuses on the risk properties and identifying optimal ridge parameters for minimizing the risks. These new risk properties reveal interesting Stein's shrinkage effects in 
thresholded regression, 
which was lacking before.

\section{Implementation} \label{Sec5}
Efficient algorithms for the implementation of regularization methods include the LQA (Fan and Li, 2001), LARS (Efron et al., 2004), and LLA (Zou and Li, 2008). As an alternative to these algorithms, the coordinate optimization 
has become popular due to its scalability for large-scale problems; see, for example, Friedman et al. (2007), Wu and Lange (2008), and Fan and Lv (2011).
In this paper, we apply the ICA algorithm (Fan and Lv, 2011) to implement the regularization methods. 
See Section V in Fan and Lv (2011) for a detailed description of this algorithm. An analysis of convergence properties of this algorithm has been presented in Lin and Lv (2013). In particular, the univariate global minimizer for each coordinate admits a closed form as given in Lemma \ref{L1}, for both hard-thresholding penalty $p_{H, \lambda}(t)$ and $L_0$-penalty $p_{H_0, \lambda}(t)$. We would like to point out that the algorithm is not guaranteed to find the global minimizer.

Although our theory relies on the union of coordinate subspaces $\mathbb{S}_{M/2}$ associated with the robust spark of the design matrix, the implementation via the ICA algorithm does not require the knowledge of such a space. It is a path-following algorithm, based on a decreasing grid of regularization parameter $\lambda$, that produces a sequence of most sparse solutions to less sparse solutions, with the solution given by the previous $\lambda$ as a warm start for the next $\lambda$. The collinearity of sparse models can be tracked easily by calculating the smallest singular value of the subdesign matrix given by the support of each produced sparse solution.

To better illustrate our theoretical results and make a fair comparison of all methods, we select the tuning parameters by minimizing the prediction error calculated using an independent validation set, with size equal to the sample size in the simulation study. We use the SICA (Lv and Fan, 2009) with penalty $p_\lambda(t; a) = \lambda (a + 1)t/(a + t)$, with a small shape parameter $a$ such as $10^{-4}$ or $10^{-2}$, as a proxy of the $L_0$-regularization method. Following Lin and Lv (2013), some pilot solutions with larger values of $a$ are computed to stabilize the solution. See also Lin and Lv (2013) for the closed-form solution of the univariate SICA estimator.

\section{Numerical studies} \label{Sec6}
In this section, we investigate the finite-sample performance of regularization methods with hard-thresholding (Hard) and SICA penalties, with comparison to the Lasso and oracle procedure which knew the true model in advance. We consider both cases of light-tailed and heavy-tailed errors, with Gaussian distribution for the former and $t$-distribution for the latter.

\subsection{Simulation examples} \label{Sec6.1}

\subsubsection{Simulation example 1} \label{Sec6.1.1}
We first consider the linear regression model (\ref{001}) with Gaussian error $\bveps \sim N(\textbf{0},\sigma^2 I_n)$. We generated 100 data sets from this model with true regression coefficient vector $\bbeta_0 = (\bv^T,\cdots,\bv^T,\bzero\t)^T$ with the pattern $\bv = (\bbeta_{\text{strong}}^T, \bbeta_{\text{weak}}^T)^T$ repeated $q$ times, where $\bbeta_{\text{strong}} = (0.6, 0, 0, -0.6, 0, 0)\t$ and  $\bbeta_{\text{weak}} = (0.05, 0, 0, -0.05, 0, 0)\t$ or $(0.1, 0, 0, -0.1, 0, 0)\t$. The coefficient subvectors $\bbeta_{\text{strong}}$ and $\bbeta_{\text{weak}}$ stand for the strong signals and weak signals in $\bbeta_0$, respectively. The two choices of $\bbeta_{\text{weak}}$ showed the performance of four methods under different levels of weak signals. We set $q = 3$ so that there are six strong signals (with magnitude 0.6) and six weak signals (with magnitude 0.05 or 0.1) in the true coefficient vector. The sample size $n$ was chosen to be $100$ and two settings of $(p, \sigma) = (1000, 0.4)$ and $(5000, 0.3)$ were considered. For each data set, all the rows of the $n \times p$ design matrix $\bX$ were sampled as independent and identically distributed 
copies from a multivariate normal distribution $N(\bzero, \Sig)$ with $\Sig = (0.5^{|i-j|})_{1 \leq i, j \leq p}$. This allows for correlation among the covariates at the population level. The sample collinearity among the covariates can be at an even higher level due to high dimensionality. We applied the Lasso, Hard, and SICA to produce a sequence of sparse models and selected the tuning parameters as discussed in Section \ref{Sec5}.

\begin{table}
\caption{\label{Tab1} Means and standard deviations (in parentheses) of different performance measures by all methods over 100 simulations in Section \ref{Sec6.1.1}}
\centering
\vspace{0.1in}
\fbox{%
\begin{tabular}{llcccc}
Setting    &         Measure            &    Lasso             &     Hard             &   SICA            &    Oracle \\
\hline
$p = 1000$       &         PE                 &    0.3025 (0.0479)   &     0.1862 (0.0086)  &   0.1862 (0.0103) &    0.1829 (0.0100) \\
$|\beta_{\text{weak}}| = 0.05$        &         $L_2$-loss         &    0.4007 (0.0653)   &     0.1679 (0.0238)  &   0.1678 (0.0276) &    0.1505 (0.0324) \\
       &         $L_1$-loss         &    1.7660 (0.2942)   &     0.5274 (0.0769)  &   0.5276 (0.0921) &    0.4277 (0.0979) \\
           &         $L_{\infty}$-loss  &    0.2012 (0.0418)   &     0.0804 (0.0258)  &   0.0790 (0.0255) &    0.0854 (0.0207) \\
           &         FP                 &    33.7900 (7.0457)  &     0.0800 (0.2727)  &   0.0900 (0.4044) &    0 (0) \\
           &         FN-strong          &    0 (0)             &     0 (0)            &   0 (0)           &    0 (0) \\
           &         FN-weak            &    5.6000 (0.6513)   &     5.9900 (0.1000)  &   5.9900 (0.1000) &    0 (0) \\
           &         $\widehat{\sigma}$ &    0.4295 (0.0473)   &     0.4158 (0.0328)  &   0.4155 (0.0351) &    0.4000 (0.0347) \\
\cline{2-6}
$p = 1000$      &         PE                 &    0.3643 (0.0584)   &     0.2272 (0.0115)  &   0.2283 (0.0124) &    0.1829 (0.0100) \\
$|\beta_{\text{weak}}| = 0.1$       &         $L_2$-loss         &    0.4882 (0.0674)   &     0.2749 (0.0223)  &   0.2769 (0.0224) &    0.1505 (0.0324) \\
        &         $L_1$-loss         &    2.2134 (0.3202)   &     0.8466 (0.1018)  &   0.8553 (0.1052) &    0.4277 (0.0979) \\
           &         $L_{\infty}$-loss  &    0.2225 (0.0453)   &     0.1068 (0.0177)  &   0.1077 (0.0177) &    0.0854 (0.0207) \\
           &         FP                 &    34.4300 (6.9866)  &     0.0900 (0.3208)  &   0.1600 (0.5453) &    0 (0) \\
           &         FN-strong          &    0 (0)             &     0 (0)            &   0 (0)           &    0 (0) \\
           &         FN-weak            &    4.9200 (0.9711)   &     5.8200 (0.6257)  &   5.8000 (0.5125) &    0 (0) \\
           &         $\widehat{\sigma}$ &    0.4676 (0.0541)   &     0.4559 (0.0377)  &   0.4540 (0.0425) &    0.4000 (0.0347) \\
\cline{2-6}
$p = 5000$       &         PE                 &    0.2634 (0.0744)   &     0.1097 (0.0058)  &   0.1088 (0.0039) &    0.1027 (0.0062) \\
$|\beta_{\text{weak}}| = 0.05$        &         $L_2$-loss         &    0.4419 (0.0904)   &     0.1476 (0.0185)  &   0.1450 (0.0127) &    0.1122 (0.0260) \\
       &         $L_1$-loss         &    1.8507 (0.3387)   &     0.4593 (0.0602)  &   0.4528 (0.0464) &    0.3166 (0.0775) \\
           &         $L_{\infty}$-loss  &    0.2188 (0.0507)   &     0.0621 (0.0206)  &   0.0592 (0.0152) &    0.0663 (0.0188) \\
           &         FP                 &    37.3900 (4.9826)  &     0.0600 (0.2778)  &   0.0100 (0.1000) &    0 (0) \\
           &         FN-strong          &    0 (0)             &     0 (0)            &   0 (0)           &    0 (0) \\
           &         FN-weak            &    5.8600 (0.3487)   &     5.9900 (0.1000)  &   6.0000 (0)      &    0 (0) \\
           &         $\widehat{\sigma}$           &    0.3822 (0.0452)   &     0.3173 (0.0239)  &   0.3187 (0.0231) &    0.2976 (0.0242) \\
\cline{2-6}
$p = 5000$       &         PE                 &    0.3603 (0.1089)   &     0.1838 (0.2401)  &   0.1489 (0.0070) &    0.1027 (0.0062) \\
$|\beta_{\text{weak}}| = 0.1$        &         $L_2$-loss         &    0.5594 (0.1054)   &     0.2830 (0.1654)  &   0.2581 (0.0136) &    0.1122 (0.0260) \\
        &         $L_1$-loss         &    2.4396 (0.3980)   &     0.8361 (0.4546)  &   0.7685 (0.1077) &    0.3166 (0.0775) \\
           &         $L_{\infty}$-loss  &    0.2584 (0.0618)   &     0.1117 (0.0704)  &   0.1016 (0.0066) &    0.0663 (0.0188) \\
           &         FP                 &    38.6000 (4.2593)  &     0.0700 (0.4324)  &   0.2200 (1.8123) &    0 (0) \\
           &         FN-strong          &    0 (0)             &     0.1100 (0.7771)  &   0 (0)           &    0 (0) \\
           &         FN-weak            &    5.5300 (0.6269)   &     5.7700 (0.5096)  &   5.7100 (0.6403) &    0 (0) \\
           &         $\widehat{\sigma}$           &    0.4417 (0.0557)   &     0.3826 (0.1214)  &   0.3629 (0.0429) &    0.2976 (0.0242) \\
\end{tabular}}
\end{table}

\begin{table}
\caption{\label{Tab2} Means and standard deviations (in parentheses) of different performance measures by all methods followed by the $L_2$-regularization over 100 simulations in Section \ref{Sec6.1.1}}
\centering
\vspace{0.1in}
\fbox{%
\begin{tabular}{llccccc}
Setting   &         Measure           &      Lasso-$L_2$        &  Hard-$L_2$         &  SICA-$L_2$         &   Oracle-$L_2$ \\
\hline
$p = 1000$      &         PE                &      0.3501 (0.0538)    &  0.1851 (0.0083)    &  0.1852 (0.0101)    &   0.1812 (0.0098) \\
$|\beta_{\text{weak}}| = 0.05$       &         $L_2$-loss        &      0.4464 (0.0635)    &  0.1658 (0.0237)    &  0.1657 (0.0277)    &   0.1437 (0.0329) \\
      &         $L_1$-loss        &      2.5473 (0.3986)    &  0.5169 (0.0764)    &  0.5177 (0.0928)    &   0.4061 (0.1001) \\
          &         $L_{\infty}$-loss &      0.1698 (0.0412)    &  0.0752 (0.0250)    &  0.0741 (0.0248)    &   0.0772 (0.0206) \\
\cline{2-6}
$p = 1000$      &         PE                &      0.4168 (0.0675)    &  0.2257 (0.0109)    &  0.2270 (0.0117)    &   0.1812 (0.0098) \\
$|\beta_{\text{weak}}| = 0.1$        &         $L_2$-loss        &      0.5272 (0.0688)    &  0.2734 (0.0221)    &  0.2755 (0.0218)    &   0.1435 (0.0328) \\
       &         $L_1$-loss        &      3.0270 (0.4430)    &  0.8366 (0.1016)    &  0.8450 (0.1045)    &   0.4053 (0.0990) \\
          &         $L_{\infty}$-loss &      0.1905 (0.0472)    &  0.1053 (0.0150)    &  0.1060 (0.0142)    &   0.0770 (0.0204) \\
\cline{2-6}
$p = 5000$      &         PE                &      0.2642 (0.0532)    &  0.1090 (0.0055)    &  0.1082 (0.0037)    &   0.1020 (0.0060) \\
$|\beta_{\text{weak}}| = 0.05$        &         $L_2$-loss        &      0.4358 (0.0675)    &  0.1462 (0.0181)    &  0.1437 (0.0127)    &   0.1082 (0.0263) \\
      &         $L_1$-loss        &      2.4131 (0.3343)    &  0.4508 (0.0586)    &  0.4448 (0.0462)    &   0.3019 (0.0758) \\
          &         $L_{\infty}$-loss &      0.1896 (0.0454)    &  0.0597 (0.0195)    &  0.0569 (0.0146)    &   0.0610 (0.0197) \\
\cline{2-6}
$p = 5000$      &         PE                &      0.3594 (0.0816)    &  0.1830 (0.2403)    &  0.1481 (0.0071)    &   0.1020 (0.0060) \\
$|\beta_{\text{weak}}| = 0.1$        &         $L_2$-loss        &      0.5492 (0.0833)    &  0.2823 (0.1656)    &  0.2574 (0.0141)    &   0.1082 (0.0264) \\
       &         $L_1$-loss        &      3.0841 (0.4236)    &  0.8280 (0.4560)    &  0.7614 (0.1090)    &   0.3017 (0.0760) \\
          &         $L_{\infty}$-loss &      0.2242 (0.0562)    &  0.1116 (0.0704)    &  0.1013 (0.0058)    &   0.0610 (0.0198) \\
\end{tabular}}
\end{table}

\begin{figure}
\centering
\makebox{\hspace{-0in}\includegraphics[scale=0.75]%
{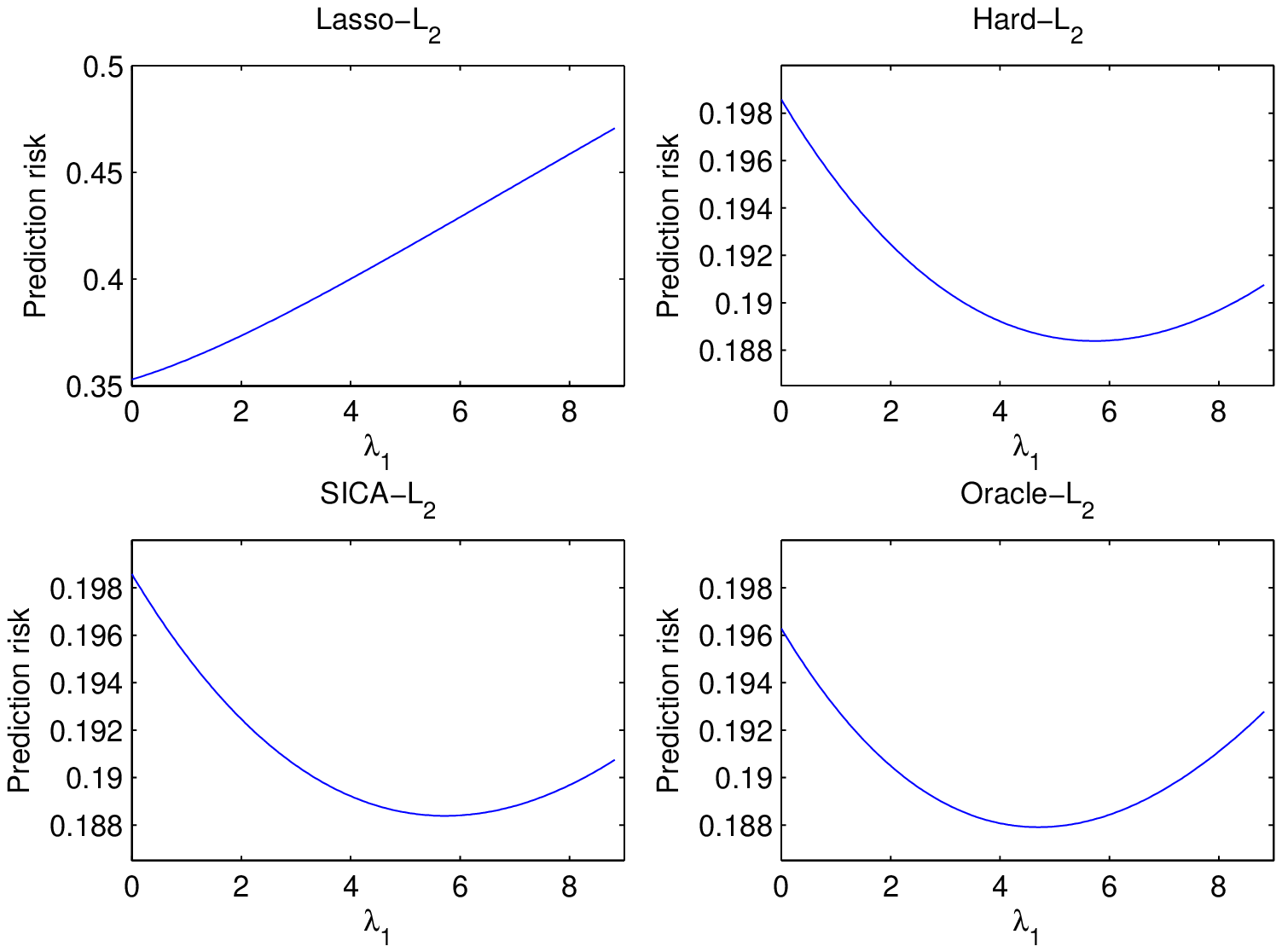}%
}%
\vspace{-0.2in}
\caption{Representative prediction risk curves as a function of the ridge parameter $\lambda_1$ by all methods in Section \ref{Sec6.1.1} for the case of $(p, |\beta_{\text{weak}}|) = (1000, 0.05)$.}
\label{Fig3}%
\end{figure}%

\begin{figure}
\centering
\makebox{\hspace{-0in}\includegraphics[scale=0.75]%
{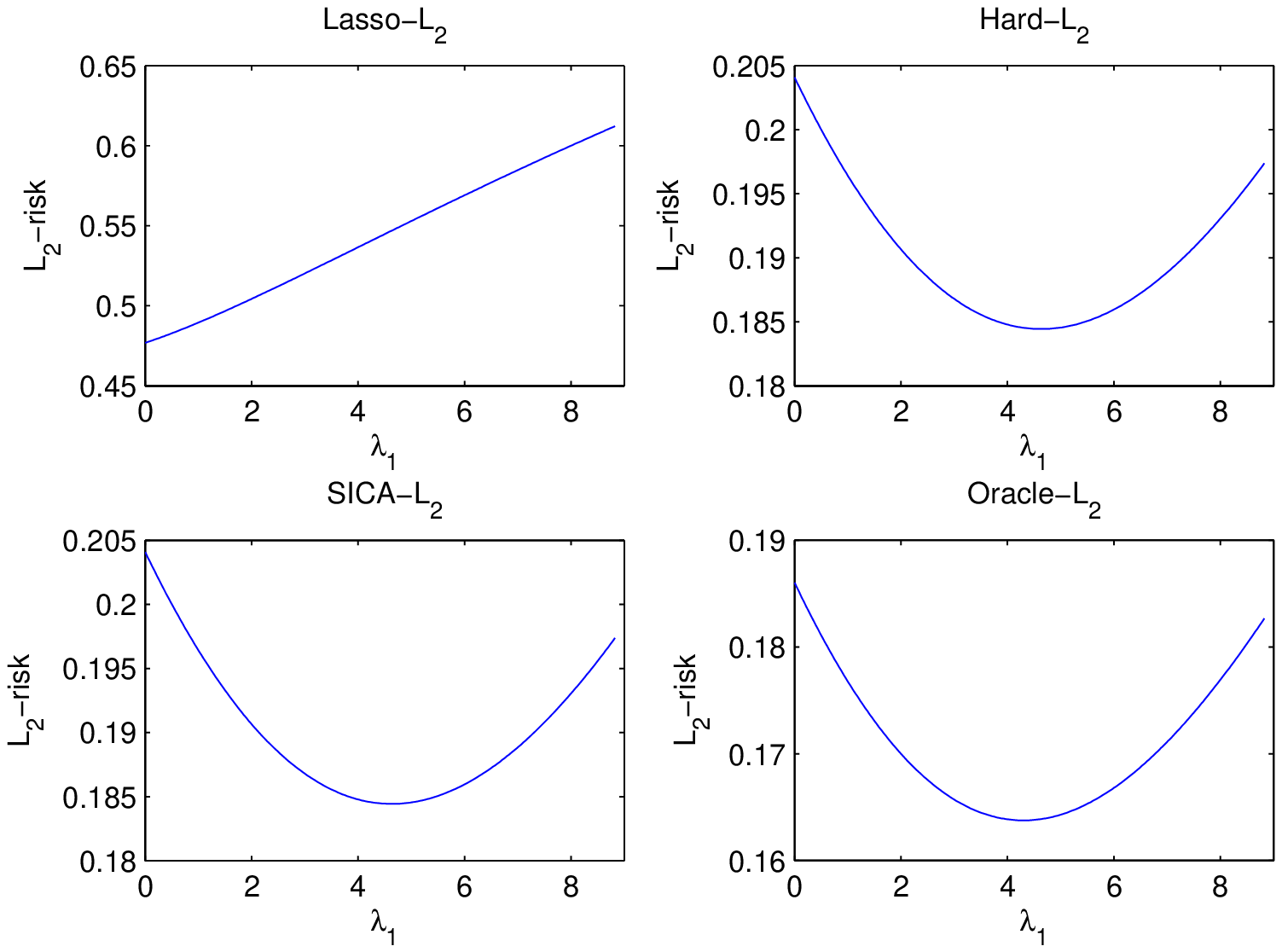}%
}%
\vspace{-0.2in}
\caption{Representative $L_2$-risk curves as a function of the ridge parameter $\lambda_1$ by all methods in Section \ref{Sec6.1.1} for the case of $(p, |\beta_{\text{weak}}|) = (1000, 0.05)$.}
\label{Fig4}%
\end{figure}%

The overall signal-to-noise ratios in the settings of $(p, |\beta_{\text{weak}}|) = (1000, 0.05)$, $(1000, 0.1)$, $(5000, 0.05)$, and $(5000, 0.1)$ are 11.70, 11.77, 20.80, and 20.92, respectively. These overall measures, however, do not reflect the individual signal strength for each strong or weak signal, which measures the difficulty of the variable selection problem. In the case of $p = 1000$, the individual signal-to-noise ratio is $0.6^2/\sigma^2 = 2.25$ for each strong signal, and $0.05^2/\sigma^2 = 0.0156$ or $0.1^2/\sigma^2 = 0.0625$ for each weak signal with level $0.05$ or $0.1$. In the case of $p = 5000$, the individual signal-to-noise ratio is $0.6^2/\sigma^2 = 4$ for each strong signal, and $0.05^2/\sigma^2 = 0.0278$ or $0.1^2/\sigma^2 = 0.1111$ for each weak signal with level $0.05$ or $0.1$. We see that the six weak covariates have very low signal strength. Their signal strength is even lower when the high dimensionality is taken into account, due to the well-known phenomenon of noise accumulation in high dimensions.

To compare the three regularization methods with the oracle procedure, we consider several performance measures. The first measure is the prediction error (PE) defined as $E (Y - \bx\t \hbbeta)^2$ with $\hbbeta$ an estimate and $(\bx\t, Y)$ an independent observation of the covariates and response. To calculate the expectation, we generated an independent test sample of size $10,000$. The second to fourth measures are the $L_q$-estimation losses $\|\hbbeta - \bbeta_0\|_q$ with $q = 2, 1$, and $\infty$, respectively. The fifth to seventh measures are the number of false positives (FP), and numbers of false negatives for strong signals (FN-strong) and false negatives for weak signals (FN-weak) for variable selection, where a false positive means a falsely selected noise covariate in the model and a false negative means a missed true covariate. We also compare the estimated error standard deviation $\widehat{\sigma}$ by all methods.

Table \ref{Tab1} summarizes the comparison results by all methods. As seen in the measure of FN-weak, the weak covariates tended to be excluded by each regularization method since they have very low signal strength. At the weak signal level of $0.05$, thanks to their concavity both Hard and SICA followed very closely the oracle procedure in terms of all other measures, while the Lasso produced a much larger model with lower prediction and variable selection accuracy due to its well-known bias issue. When the weak signal level increases to $0.1$, the performance of each method deteriorated due to the difficulty of recovering weak covariates. We also considered the case of no weak signals with $\bbeta_{\text{weak}} = \bzero$. In such case, all methods performed better and their relative performance was the same as in the case with the weak signal level of $0.05$, with both Hard and SICA having almost identical performance as the oracle procedure. To save space, these additional simulation results are not included here but are available upon request.

We also investigate the risk properties and shrinkage effects of the $L_2$-regularized refitted estimators $\hbbeta_{\text{refitted}}$ defined in (\ref{013}) for all methods. Table \ref{Tab2} presents the performance of these shrinkage estimators in the above two settings with the ridge parameter $\lambda_1$ selected to minimize the corresponding risks. A comparison of risks under different losses in Tables \ref{Tab1} and \ref{Tab2} shows the improvement of the $L_2$-regularized refitted estimators over the estimators given by Hard, SICA, and oracle procedure, respectively. These numerical results are in line with the theoretical results in Theorem \ref{Thm5}. The results of the $L_2$-regularized refitted estimator for the Lasso show no improvement in risks. This is because of the bias issue of the Lasso giving rise to a large model. Figures \ref{Fig3} and \ref{Fig4} depict some representative risk curves as a function of the ridge parameter $\lambda_1$ by all methods for the prediction loss and $L_2$-loss, respectively. These plots demonstrate Stein's shrinkage effects for the thresholded regression followed by the $L_2$-regularization under both estimation and prediction risks.

\subsubsection{Simulation example 2} \label{Sec6.1.2}
A natural question is whether the results and phenomena for light-tailed errors hold for heavy-tailed errors or not. We now turn our attention to such a case for the linear regression model (\ref{001}) with $t$ error distribution. The setting of this simulation example is the same as that in Section \ref{Sec6.1.1} except that the error vector is $\bveps = \sigma \bleta$, where the components of the $n$-dimensional random vector $\bleta$ are independent and follow the $t$-distribution with $df = 10$ degrees of freedom. We compared the Lasso, Hard, and SICA with the oracle procedure in the same two settings of $(p, \sigma) = (1000, 0.4)$ and $(5000, 0.3)$. The same performance measures as in Section \ref{Sec6.1.1} are employed for comparison.

\begin{table}
\caption{\label{Tab5} Means and standard deviations (in parentheses) of different performance measures by all methods over 100 simulations in Section \ref{Sec6.1.2}; the population error standard deviation (SD) $\sigma \sqrt{df/(df-2)}$ equals $0.4472$ in the case of $p = 1000$, and $0.3354$ in the case of $p = 5000$}
\centering
\vspace{0.1in}
\fbox{%
\begin{tabular}{llcccc}
Setting    &         Measure            &    Lasso             &     Hard             &   SICA            &    Oracle \\
\hline
$p = 1000$       &         PE                 &    0.3845 (0.0705)   &     0.2277 (0.0137)  &   0.2285 (0.0168) &    0.2276 (0.0151) \\
$|\beta_{\text{weak}}| = 0.05$         &         $L_2$-loss         &    0.4547 (0.0801)   &     0.1718 (0.0316)  &   0.1734 (0.0361) &    0.1655 (0.0415) \\
       &         $L_1$-loss         &    1.9683 (0.3523)   &     0.5335 (0.0932)  &   0.5386 (0.1124) &    0.4682 (0.1202) \\
           &         $L_{\infty}$-loss  &    0.2306 (0.0554)   &     0.0858 (0.0347)  &   0.0870 (0.0360) &    0.0937 (0.0283) \\
           &         FP                 &    32.7800 (8.6311)  &     0.0600 (0.2778)  &   0.1000 (0.4606) &    0 (0) \\
           &         FN-strong          &    0 (0)             &     0 (0)            &   0 (0)           &    0 (0) \\
           &         FN-weak            &    5.6200 (0.6321)   &     6.0000 (0)       &   6.0000 (0)      &    0 (0) \\
           &         Error SD           &    0.4867 (0.0599)   &     0.4652 (0.0412)  &   0.4645 (0.0417) &    0.4517 (0.0368) \\
\cline{2-6}
$p = 1000$       &         PE                 &    0.4462 (0.0777)   &     0.2693 (0.0149)  &   0.2702 (0.0172) &    0.2276 (0.0151) \\
$|\beta_{\text{weak}}| = 0.1$         &         $L_2$-loss         &    0.5331 (0.0773)   &     0.2787 (0.0245)  &   0.2797 (0.0272) &    0.1655 (0.0415) \\
        &         $L_1$-loss         &    2.4177 (0.3695)   &     0.8557 (0.1003)  &   0.8628 (0.1188) &    0.4682 (0.1202) \\
           &         $L_{\infty}$-loss  &    0.2491 (0.0558)   &     0.1123 (0.0250)  &   0.1131 (0.0259) &    0.0937 (0.0283) \\
           &         FP                 &    34.1400 (8.1996)  &     0.0600 (0.2387)  &   0.1300 (0.6139) &    0 (0) \\
           &         FN-strong          &    0 (0)             &     0 (0)            &   0 (0)           &    0 (0) \\
           &         FN-weak            &    5.0200 (0.9209)   &     5.9200 (0.2727)  &   5.8600 (0.4499) &    0 (0) \\
           &         Error SD           &    0.5152 (0.0633)   &     0.5033 (0.0445)  &   0.5004 (0.0512) &    0.4517 (0.0368) \\
\cline{2-6}
$p = 5000$       &         PE                 &    0.3295 (0.1096)   &     0.1343 (0.0058)  &   0.1343 (0.0060) &    0.1277 (0.0068) \\
$|\beta_{\text{weak}}| = 0.05$         &         $L_2$-loss         &    0.4897 (0.1151)   &     0.1539 (0.0168)  &   0.1541 (0.0169) &    0.1226 (0.0270) \\
       &         $L_1$-loss         &    2.0497 (0.4196)   &     0.4831 (0.0588)  &   0.4833 (0.0586) &    0.3411 (0.0796) \\
           &         $L_{\infty}$-loss  &    0.2439 (0.0625)   &     0.0684 (0.0209)  &   0.0688 (0.0212) &    0.0717 (0.0191) \\
           &         FP                 &    38.4600 (5.4558)  &     0.0300 (0.1714)  &   0.0300 (0.1714) &    0 (0) \\
           &         FN-strong          &    0 (0)             &     0 (0)            &   0 (0)           &    0 (0) \\
           &         FN-weak            &    5.8800 (0.3266)   &     5.9800 (0.1407)  &   5.9800 (0.1407) &    0 (0) \\
           &         Error SD           &    0.4254 (0.0569)   &     0.3560 (0.0319)  &   0.3560 (0.0319) &    0.3366 (0.0321) \\
\cline{2-6}
$p = 5000$       &         PE                 &    0.4307 (0.1419)   &     0.1761 (0.0080)  &   0.1767 (0.0132) &    0.1277 (0.0068) \\
$|\beta_{\text{weak}}| = 0.1$         &         $L_2$-loss         &    0.6030 (0.1245)   &     0.2671 (0.0146)  &   0.2680 (0.0219) &    0.1226 (0.0270) \\
        &         $L_1$-loss         &    2.6203 (0.4894)   &     0.8068 (0.0701)  &   0.8150 (0.1238) &    0.3411 (0.0796) \\
           &         $L_{\infty}$-loss  &    0.2845 (0.0722)   &     0.1047 (0.0143)  &   0.1035 (0.0105) &    0.0717 (0.0191) \\
           &         FP                 &    38.3900 (5.0510)  &     0.0500 (0.2190)  &   0.1800 (1.2092) &    0 (0) \\
           &         FN-strong          &    0 (0)             &     0 (0)            &   0 (0)           &    0 (0) \\
           &         FN-weak            &    5.5900 (0.5702)   &     5.8500 (0.3860)  &   5.8100 (0.4648) &    0 (0) \\
           &         Error SD           &    0.4828 (0.0625)   &     0.4039 (0.0354)  &   0.4005 (0.0458) &    0.3366 (0.0321) \\
\end{tabular}}
\end{table}

\begin{table}
\caption{\label{Tab6} Means and standard deviations (in parentheses) of different performance measures by all methods followed by the $L_2$-regularization over 100 simulations in Section \ref{Sec6.1.2}}
\centering
\vspace{0.1in}
\fbox{%
\begin{tabular}{llccccc}
Setting   &         Measure           &      Lasso-$L_2$        &  Hard-$L_2$         &  SICA-$L_2$         &   Oracle-$L_2$ \\
\hline
$p = 1000$      &         PE                &      0.4447 (0.0747)    &  0.2263 (0.0135)    &  0.2270 (0.0146)    &   0.2256 (0.0148) \\
$|\beta_{\text{weak}}| = 0.05$        &         $L_2$-loss        &      0.5059 (0.0767)    &  0.1686 (0.0321)    &  0.1701 (0.0342)    &   0.1588 (0.0411) \\
      &         $L_1$-loss        &      2.8356 (0.5052)    &  0.5191 (0.0929)    &  0.5238 (0.1013)    &   0.4426 (0.1157) \\
          &         $L_{\infty}$-loss &      0.1976 (0.0541)    &  0.0796 (0.0328)    &  0.0808 (0.0335)    &   0.0858 (0.0280) \\
\cline{2-6}
$p = 1000$      &         PE                &      0.5170 (0.0835)    &  0.2676 (0.0149)    &  0.2684 (0.0172)    &   0.2256 (0.0148) \\
$|\beta_{\text{weak}}| = 0.1$        &         $L_2$-loss        &      0.5828 (0.0767)    &  0.2770 (0.0250)    &  0.2780 (0.0277)    &   0.1588 (0.0412) \\
       &         $L_1$-loss        &      3.3344 (0.5248)    &  0.8426 (0.1047)    &  0.8491 (0.1230)    &   0.4427 (0.1156) \\
          &         $L_{\infty}$-loss &      0.2180 (0.0567)    &  0.1099 (0.0218)    &  0.1107 (0.0227)    &   0.0858 (0.0281) \\
\cline{2-6}
$p = 5000$      &         PE                &      0.3312 (0.0815)    &  0.1335 (0.0055)    &  0.1335 (0.0056)    &   0.1268 (0.0066) \\
$|\beta_{\text{weak}}| = 0.05$        &         $L_2$-loss        &      0.4858 (0.0877)    &  0.1520 (0.0165)    &  0.1522 (0.0165)    &   0.1180 (0.0275) \\
      &         $L_1$-loss        &      2.6985 (0.4105)    &  0.4719 (0.0587)    &  0.4724 (0.0581)    &   0.3256 (0.0807) \\
          &         $L_{\infty}$-loss &      0.2090 (0.0523)    &  0.0645 (0.0182)    &  0.0649 (0.0187)    &   0.0653 (0.0182) \\
\cline{2-6}
$p = 5000$      &         PE                &      0.4332 (0.1092)    &  0.1750 (0.0075)    &  0.1757 (0.0132)    &   0.1268 (0.0066) \\
$|\beta_{\text{weak}}| = 0.1$       &         $L_2$-loss        &      0.5954 (0.0995)    &  0.2659 (0.0144)    &  0.2668 (0.0221)    &   0.1178 (0.0275) \\
      &         $L_1$-loss        &      3.3325 (0.5086)    &  0.7961 (0.0715)    &  0.8053 (0.1260)    &   0.3254 (0.0808) \\
          &         $L_{\infty}$-loss &      0.2467 (0.0623)    &  0.1036 (0.0113)    &  0.1029 (0.0100)    &   0.0649 (0.0180) \\
\end{tabular}}
\end{table}

The means and standard deviations of different performance measures by all methods are listed in Table \ref{Tab5}. Table \ref{Tab6} details the performance of the $L_2$-regularized refitted estimators, as described in Section \ref{Sec6.1.1}, 
with the ridge parameter $\lambda_1$ selected to minimize the corresponding risks. The conclusions are similar to those in Section \ref{Sec6.1.1}. By comparing the results in this simulation example with those in Section \ref{Sec6.1.1} for Gaussian error, we see that the performance of all methods deteriorated when the error distribution becomes heavy-tailed. Both Hard and SICA still followed closely the oracle procedure at the weak signal level of $0.05$. We also observe the phenomenon of Stein's shrinkage effects for the thresholded regression followed by the $L_2$-regularization under both estimation and prediction risks in this case of heavy-tailed error distribution.

\subsection{Real data example} \label{Sec6.2}
We apply the Lasso, Hard, and SICA, as well as these methods followed by the $L_2$-regularization, to the diabetes data set studied in Efron et al. (2004). This data set consists of measurements for $n = 442$ diabetes patients on the response variable, a quantitative measure of disease progression one year after baseline, and ten baseline variables: sex (\verb"sex"), age (\verb"age"), body mass index (\verb"bmi"), average blood pressure (\verb"bp"), and six blood serum measurements (\verb"tc", \verb"ldl", \verb"hdl", \verb"tch", \verb"ltg", \verb"glu"). Efron et al. (2004) considered the quadratic model with interactions, by adding the squares of all baseline variables except the dummy variable \verb"sex", and all interactions between each pair of the ten baseline variables. This results in a linear regression model with $p = 64$ predictors. We adopt this model to analyze the diabetes data set.

\begin{figure}
\centering
\makebox{\hspace{-0.15in}\includegraphics[scale=0.75]%
{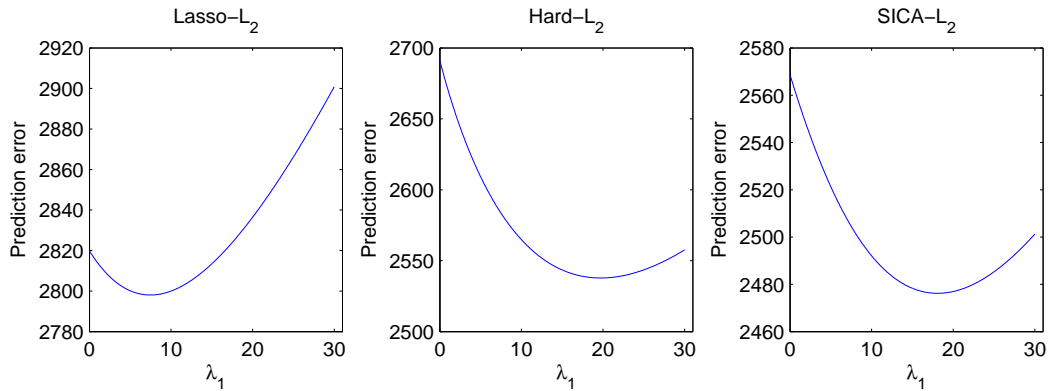}%
}%
\vspace{-0.10in}
\caption{Representative prediction error curves as a function of the ridge parameter $\lambda_1$ by all methods on the diabetes data set in Section \ref{Sec6.2}.}
\label{Fig5}%
\end{figure}%

\begin{table}
\caption{\label{Tab8} Selection probabilities ($t$-statistics, with magnitude above 2 in boldface) of most frequently selected predictors with number up to median model size
by each method across 100 random splittings of the diabetes data set in Section \ref{Sec6.2}}
\centering
\vspace{0.1in}
\fbox{%
\begin{tabular}{lccclccc}
Predictor & Lasso & Hard & SICA & Predictor & Lasso & Hard & SICA \\
\hline
sex  & 0.94 (\textbf{-2.03}) & 0.83 (\textbf{-2.16}) & 0.82 (\textbf{-2.07}) &     bp$^2$  & 0.54 (0.42) & --- & ---  \\
bmi  & 1.00 (\textbf{17.24}) & 0.99 (\textbf{6.25}) & 1.00 (\textbf{8.65}) &          glu$^2$  & 1.00 (\textbf{3.95}) & 0.50 (0.94) & 0.57 (1.10)  \\
bp  & 1.00 (\textbf{5.82}) & 0.87 (\textbf{2.51}) & 0.91 (\textbf{2.99}) &          sex$\ast$age  & 0.98 (\textbf{3.09}) & 0.87 (\textbf{2.46}) & 0.81 (\textbf{2.00}) \\
tc  & 0.43 (-0.67) & --- & --- &    sex$\ast$bp  & 0.73 (0.99) & --- & ---  \\
hdl  & 1.00 (\textbf{-3.63}) &  0.80 (-1.86) & 0.79 (-1.83) &   age$\ast$bp  & 0.87 (1.27) & --- & --- \\
ltg  & 1.00 (\textbf{9.27}) & 1.00 (\textbf{7.27}) & 1.00 (\textbf{8.22}) &          age$\ast$ltg  & 0.74 (0.82) & --- & --- \\
glu  & 0.85 (1.21) & --- & --- &   age$\ast$glu  & 0.59 (0.82) & --- & --- \\
age$^2$  & 0.94 (1.91) & --- & --- &    bmi$\ast$bp   & 0.99 (\textbf{2.25}) & 0.81 (1.94) & 0.76 (1.69) \\
bmi$^2$  & 0.98 (\textbf{2.49}) & --- & --- &   bp$\ast$hdl  & 0.47 (0.68) & --- & ---  \\
\end{tabular}}
\end{table}

We randomly split the full data set 100 times into a training set of 400 samples and a validation set of 42 samples. For each splitting of the data set, we applied each regularization method to the training set with the quadratic model, and calculated the prediction error, as defined in Section \ref{Sec6.1.1}, on the validation set. Minimizing the prediction error gives the best model for each regularization method. The means (standard deviations) of these minimum prediction errors over 100 random splittings were 2894.5 (655.5) for Lasso, 2802.5 (635.5) for Hard, and 2800.6 (615.2) for SICA. We see that both Hard and SICA improved over Lasso in prediction accuracy. The relatively large standard deviations indicate the difficulty of the prediction problem for this data set. Based on the estimated model by each method, we also investigated the $L_2$-regularized refitted estimator with ridge parameter $\lambda_1$ selected by the validation set. The means (standard deviations) of their prediction errors over 100 random splittings were 2957.9 (671.3) for Lasso-$L_2$, 2770.3 (630.9) for Hard-$L_2$, and 2770.2 (614.9) for SICA-$L_2$. We observe in Figure \ref{Fig5} shrinkage effects for both Hard and SICA followed by the $L_2$-regularization, whereas the refitting with $L_2$-regularization did not generally improve the performance of Lasso, as also shown in the simulation studies.

We also calculated the median model size by each method: 18 by Lasso, 8 by Hard, and 8 by SICA. For each method, we computed the percentage of times each predictor was selected and listed the most frequently chosen $m$ predictors in Table \ref{Tab8}, with $m$ equal to the median model size by the method. Table \ref{Tab8} also reports the $t$-statistics of selected predictors as the ratio of mean to standard deviation, with the means and standard deviations of their coefficients calculated over 100 random splittings. We see that the set of most frequently selected predictors for Hard is identical to that for SICA, which is further a subset of that for Lasso. Some of these selected predictors have $t$-statistics with magnitude below 2, indicating less significance. We also observe that the coefficients for predictors \verb"sex" and \verb"hdl" estimated by all methods are negative. It is interesting to note that the interaction term \verb"sex"$\ast$\verb"age" is found to be significant, although the predictor \verb"age" is an insignificant variable based on each method.

\section*{Acknowledgements}
We sincerely thank the Joint Editor, an Associate Editor, and a referee for their valuable comments that significantly improved the paper. This work was supported by NSF CAREER Awards DMS-0955316 and DMS-1150318 and Grants DMS-0806030 and DMS-0906784, 2010 Zumberge Individual Award from USC's James H. Zumberge Faculty Research and Innovation Fund, and USC Marshall Summer Research Funding.

\appendix

\section{Proofs of main results} \label{SecA}

\subsection{Proof of Theorem \ref{Thm4}} \label{SecA.5}

The proof contains two parts. The first part establishes the model selection consistency property of $\hbbeta$ with a suitably chosen $\lambda$. The second part proves the the oracle prediction properties using the model selection consistency property from the first part.

\medskip

\noindent \textbf{Part 1: Model selection consistency property}. We prove $\supp(\hbbeta) = \supp(\bbeta_0)$ in two steps. In the first step, it will be shown that the number of nonzero elements in $\hbbeta$ is no larger than 
$s$ conditioning on event $\mathcal{E}$ defined in (\ref{event}) (see Lemma \ref{L2} in Section \ref{SecSA.2} of Supplementary Material), when $\frac{c_2}{c}\sqrt{(2s + 1)(\log \widetilde{p})/n} < \lambda < b_0$. We prove this by using the global optimality of $\hbbeta$.

By Lemma \ref{L1} and $\lambda < b_0$, any nonzero component of the true regression coefficient vector $\bbeta_0$ or of the global minimizer $\hbbeta$ is greater than $\lambda$, which ensures that $\|p_\lambda(\hbbeta)\|_1 =  \lambda^2 \|\hbbeta\|_0/2$ and $\|p_\lambda(\bbeta_0)\|_1= s \lambda^2/2$. Thus, $\|p_\lambda(\hbbeta)\|_1-\|p_\lambda(\bbeta_0)\|_1 = (\|\hbbeta\|_0 - s)\lambda^2/2$. Denote by $\bdelta = \hbbeta - \bbeta_0$. Direct calculations yield
\begin{align}\label{simp}
\nonumber Q(\hbbeta) - Q(\bbeta_0) & = 2^{-1} \|n^{-\frac{1}{2}}\bX \bdelta\|_2^2 - n^{-1}\bveps\t \bX \bdelta + \|p_\lambda(\bbeta)\|_1-\|p_\lambda(\bbeta_0)\|_1\\
& = 2^{-1} \|n^{-\frac{1}{2}}\bX \bdelta\|_2^2 - n^{-1}\bveps\t \bX \bdelta + (\|\hbbeta\|_0 - s)\lambda^2/2.
\end{align}

On the other hand, Conditional on event $\mathcal{E}$, we have
\begin{equation}\label{E}
|n^{-1}\bveps\t \bX \bdelta| \ \leq \ \|n^{-1}\bveps\t \bX\|_{\infty}\|\bdelta\|_1 \ \leq \ c_2 \sqrt{(\log \widetilde{p})/n} \|\bdelta\|_1 \ \leq \ c_2 \sqrt{(\log \widetilde{p})/n}\|\bdelta\|_0^{\frac{1}{2}} \|\bdelta\|_2.
\end{equation}
In addition, by definition and Condition \ref{cond2}, we obtain $\|\bdelta\|_0 \leq \|\bbeta_0\|_0 + \|\hbbeta\|_0 < M$, with $M$ being the robust spark of $\bX$.  Therefore, Definition \ref{Def1} entails
\begin{equation} \label{eigen}
\|n^{-\frac{1}{2}}\bX \bdelta\|_2 \geq c \|\bdelta\|_2.
\end{equation}

Combining (\ref{simp}) with the inequalities (\ref{E}) and (\ref{eigen}) established above gives
\begin{equation} \label{Qbeta2}
Q(\hbbeta) - Q(\bbeta_0)\geq  2^{-1}c^2\|\bdelta\|_2^2 - \ c_2 \sqrt{(\log \widetilde{p})/n} \|\bdelta\|_0^{\frac{1}{2}} \|\bdelta\|_2 + (\|\hbbeta\|_0 - s) \lambda^2/2.
\end{equation}
Thus, the global optimality of $\hbbeta$ ensures that
\[2^{-1} c^2 \|\bdelta\|_2^2 - c_2 \sqrt{\frac{\log \widetilde{p}}{n}} \|\bdelta\|_0^{\frac{1}{2}} \|\bdelta\|_2 + (\|\hbbeta\|_0-s) \lambda^2/2 \leq 0.\]
Reorganizing the above inequality and collecting terms, we get
\begin{equation*}
\Big[c \|\bdelta\|_2 - \frac{c_2}{c} \sqrt{\frac{\log \widetilde{p}}{n}} \|\bdelta\|_0^{\frac{1}{2}}\Big]^2 - (\frac{c_2}{c})^2 \frac{\log \widetilde{p}}{n} \|\bdelta\|_0 + (\|\hbbeta\|_0 - s)\lambda^2 \leq 0,
\end{equation*}
which gives
\begin{equation}\label{016}
(\|\hbbeta\|_0-s)\lambda^2 \leq (\frac{c_2}{c})^2 \frac{\log \widetilde{p}}{n} \|\bdelta\|_0.
\end{equation}

We next bound the value of $\|\hbbeta\|_0$ using the above inequality (\ref{016}). Let $k=\|\hbbeta\|_0$, then $\|\bdelta\|_0 = \|\hbbeta - \bbeta_0\|_0 \leq k + s$. Thus, it follows from (\ref{016}) that
$(k-s)\lambda^2 \leq (\frac{c_2}{c})^2 \frac{\log \widetilde{p}}{n} (k + s)$. Organizing it in terms of $k$ and $s$, we get
\begin{equation} \label{equik}
k(\lambda^2 - (\frac{c_2}{c})^2 \frac{\log \widetilde{p}}{n}) \leq s(\lambda^2 + (\frac{c_2}{c})^2 \frac{\log \widetilde{p}}{n}).
\end{equation}

Since $\lambda > \frac{c_2}{c}\sqrt{(2s + 1)\log \widetilde{p}/n}$, we have $\lambda^2 - (\frac{c_2}{c})^2 (2 s + 1) \frac{\log \widetilde{p}}{n} > 0$ and $\lambda^2 c^2 n - c_2^2 \log \widetilde{p} > 2 c_2^2 s\log \widetilde{p}$. Then it follows from inequality (\ref{equik}) that
\[k \leq s \frac{(\lambda^2 + (\frac{c_2}{c})^2 \frac{\log \widetilde{p}}{n})}{(\lambda^2 - (\frac{c_2}{c})^2 \frac{\log \widetilde{p}}{n})} = s (1 + \frac{2c_2^2 \log \widetilde{p}}{\lambda^2 c^2 n - c_2^2 \log \widetilde{p}})  < s + 1.\]
Therefore, the number of nonzero elements in $\hbbeta$ satisfies $\|\hbbeta\|_0\leq s$.

The second step is based on the first step, where we will use proof by contradiction to show that $\supp(\bbeta_0) \subset \supp(\hbbeta)$ with the additional assumption $\lambda < b_0 c/\sqrt{2}$ of the theorem. Suppose that $\supp(\bbeta_0)\not\subset \supp(\hbbeta)$, then the number of missed true coefficients $k = |\supp(\bbeta_0) \backslash \supp(\hbbeta)| \geq 1$. Thus we have $\|\hbbeta\|_0 \geq s - k$ and $\|\bdelta\|_0 \leq \|\hbbeta\|_0 + \|\bbeta_0\|_0 \leq 2s$. Combining these two results with  inequality (\ref{Qbeta2}) yields
\begin{align}
Q(\hbbeta)- Q(\bbeta_0)\geq (2^{-1} c^2 \|\bdelta\|_2 - c_2\sqrt{\frac{2s\log \widetilde{p}}{n}})\|\bdelta\|_2 - k \lambda^2/2.\label{Q1}
\end{align}

Note that for each $j \in \supp(\bbeta_0) \setminus \supp(\hbbeta)$, we have $|\delta_j| = |\beta_{0,j}| \geq b_0$ with $b_0$ being the lowest signal strength in Condition \ref{cond2}. Thus, $\|\bdelta\|_2 \geq \sqrt{k}b_0$, which together with Condition \ref{cond2} entails
\[4^{-1} c^2 \|\bdelta\|_2 \geq 4^{-1} c^2 \sqrt{k}b_0 \geq 4^{-1} c^2 b_0 > c_2\sqrt{(2s\log \widetilde{p})/n}.\]
Thus, it follows from (\ref{Q1}) that
\begin{align*}
Q(\hbbeta) - Q(\bbeta_0)\geq 4^{-1} c^2 \|\bdelta\|_2^2 - k \lambda^2/2 \geq  4^{-1} c^2 k b_0^2 - k \lambda^2/2 > 0,
\end{align*}
where the last step is because of the additional assumption $\lambda < b_0 c/\sqrt{2}$. The above inequality contradicts with the global optimality of $\hbbeta$. Thus, we have $\supp(\bbeta_0) \subset \supp(\hbbeta)$. Combining this with $\|\hbbeta\|_0 \leq s$ from the first step, we know that $\supp(\hbbeta) = \supp(\bbeta_0)$.

It follows from Lemma \ref{L1} and the characterization of the penalized least-squares estimator in Theorem 1 in Lv and Fan (2009) that the hard-thresholded estimator $\hbbeta$ on its support $\supp(\hbbeta)$ is exactly the ordinary least-squares estimator constructed using covariates in $\supp(\hbbeta)$. With the model selection consistency property proved above, we have the explicit form of $\hbbeta$ on its support as $(\bX_0\t \bX_0)^{-1}\bX_0\t \by$, 
where $\bX_0$ is the submatrix of the design matrix $\bX$ consisting of columns in $\supp(\bbeta_0)$. Now we derive bounds for the prediction and estimation losses of $\hbbeta$.

\medskip

\noindent \textbf{Part 2: Prediction and estimation losses}. The idea is to get the $L_2$-estimation loss bound by the global optimality of $\hbbeta$, conditional on the event $ \mathcal{E}_1 = \mathcal{E} \cap \mathcal{E}'$ with $\mathcal{E}$ and $\mathcal{E'}$ defined in (\ref{event}) (see Lemma \ref{L2} in Section \ref{SecSA.2} of Supplementary Material).

Conditional on $\mathcal{E}_1$, we have $\|\bdelta\|_0 \leq s$ by the model selection consistency property proved above. Thus, by the Cauchy-Schwarz inequality we have
\begin{equation} \label{018}
|n^{-1}\bveps\t \bX_0 \bdelta| \ \leq \ \|n^{-1}\bveps\t \bX_0\|_{\infty}\|\bdelta\|_1 \ \leq \ c'_2 \sqrt{\frac{\log n}{n}} \|\bdelta\|_1 \ \leq \ c'_2 \sqrt{\frac{s \log n}{n}} \|\bdelta\|_2,
\end{equation}
Since (\ref{simp}) and (\ref{eigen}) are still true as they depend only on 
Condition \ref{cond2} and Definition \ref{Def1}, it follows from (\ref{018}) and the model selection consistency property $\|\hbbeta\|_0 = s$ that
\begin{align*}
&Q(\hbbeta) - Q(\bbeta_0) =  2^{-1} \|n^{-1}\bX\bdelta\|_2^2 - n^{-1}\bveps\t \bX \bdelta + (\|\hbbeta\|_0 - s)\lambda^2/2\\
& \geq  2^{-1}c^2 \|\bdelta\|_2^2 - n^{-1}\bveps\t \bX \bdelta \geq  (2^{-1} c^2 \|\bdelta\|_2 - c'_2\sqrt{\frac{s\log n}{n}})\|\bdelta\|_2.
\end{align*}

Then it follows from the global optimality of $\hbbeta$ that $2^{-1} c^2 \|\bdelta\|_2 - c'_2\sqrt{\frac{s\log n}{n}} \leq 0$, which gives the $L_2$ and $L_\infty$ estimation bound as
\begin{align*}
&\|\hbbeta - \bbeta_0\|_2 = \|\bdelta\|_2 \leq \ 2c^{-2}c'_2 \sqrt{(s\log n)/n}, \\
&\|\hbbeta-\bbeta_0\|_{\infty} \leq \|\hbbeta-\bbeta_0\|_2 \leq 2c^{-2}c'_2 \sqrt{(s\log n)/n}.
\end{align*}

For $L_q$-estimation loss with $1 \leq q < 2$, applying H\"{o}lder's inequality gives
\begin{align} \label{019}
\nonumber \|\hbbeta-\bbeta_0\|_q& = (\sum_{j=1}^n |\delta_j|^q)^{1/q} \leq (\sum_{j=1}^n|\delta_j|^2)^{\frac{1}{2}}(\sum_{\delta_j \neq 0} 1^{\frac{2}{2-q}})^{\frac{1}{q}-\frac{1}{2}} = \|\bdelta\|_2 \|\bdelta\|_0^{\frac{1}{q}-\frac{1}{2}}\\
 &\leq  2c^{-2}c_2' s^{\frac{1}{q}} \sqrt{(\log n)/n}.
\end{align}

Finally we prove the bound for oracle prediction loss. Since $\hbbeta$ is the global minimizer, it follows from (\ref{simp})  and the model selection consistency property that conditioning on $\mathcal{E}_1$
\begin{align*}
& 2^{-1/2}n^{-\frac{1}{2}}\|\bX (\hbbeta-\bbeta_0)\|_2 \leq \left\{n^{-1}\bveps\t \bX \bdelta - (\|\hbbeta\|_0 - s)\lambda^2/2 \right\}^{1/2} \\
& \leq \left\{\|n^{-1}\bX\t_0\bveps\|_\infty \|\bdelta\|_1\right\}^{1/2} \leq c'_2c^{-1} \sqrt{2s (\log n)/n},
\end{align*}
where the last step is because of the $L_1$ estimation bound proved above. This completes the proof.

\subsection{Proof of Theorem \ref{Thm5}} \label{SecA.6}
In this proof, we apply mathematical techniques such as singular value decomposition and Taylor expansion to study the explicit forms of risks of the refitted estimator $\hbbeta_{\text{refitted}}$ under squared $L_2$-loss and squared prediction loss, and to find out the orders and leading terms of the optimal tuning parameter $\lambda_1$ and the corresponding minimized risks. The proof consists of two parts.

\medskip

\noindent \textbf{Part 1: Risk properties for $\hbbeta_{\text{refitted}}$ under $L_q$-estimation loss}. We first consider the risk of $\hbbeta_{\text{refitted}}$ under the squared $L_2$-loss and find the order and leading term of the corresponding optimal $\lambda_1$. The main idea is to divide the risk into two parts, and then minimize the first part conditional on event $\mathcal{E}$ defined in (\ref{event}), and show that the other part has a smaller order. 
By default, all arguments below are conditional on $\mathcal{E}$. 

Proof of Theorem \ref{Thm4} ensures that $\supp(\hbbeta) = \supp(\bbeta_0)$ conditional on event $\mathcal{E}$ under Conditions \ref{cond1} and \ref{cond2}.
Thus, if we denote $\bX_0$ as the oracle design matrix, then $\bX_1 = \bX_0$ and $s_1 = s_0$. Let $I_s$ be the $s\times s$ identity matrix for a positive integer $s$. It follows that
\[\hbbeta_{\text{refitted}} = (\bX_1\t \bX_1 + \lambda_1 I_{s_1})^{-1} \bX_1\t \by 
= (\bX_0\t \bX_0 + \lambda_1 I_s)^{-1} \bX_0\t \bX_0 \bbeta_0 + (\bX_0\t \bX_0 + \lambda_1 I_s)^{-1} \bX_0\t\bveps,\]
where in the last step we used $\by = \bX_0 \bbeta_0 + \bveps$.
So the difference between $\hbbeta_{\text{refitted}}$ and $\bbeta_0$ is
\[\hbbeta_{\text{refitted}} - \bbeta_0 = - \lambda_1(\bX_0\t \bX_0 + \lambda_1 I_s)^{-1} \bbeta_0 + (\bX_0\t \bX_0 + \lambda_1 I_s)^{-1} \bX_0\t\bveps.\]
Set $\bmu = - \lambda_1(\bX_0\t \bX_0 + \lambda_1 I_s)^{-1} \bbeta_0$ and $A = \bX_0 (\bX_0\t \bX_0 + \lambda_1 I_s)^{-1}$, then $\hbbeta_{\text{refitted}} - \bbeta_0 = \bmu + A\t \bveps$. Thus, conditioning on $\mathcal{E}$ we have
\begin{align}\label{017}
\|\hbbeta_{\text{refitted}} - \bbeta_0\|_2^2 = \bmu\t\bmu + 2\bmu\t A\t \bveps + \bveps\t AA\t \bveps.
\end{align}

In view of (\ref{017}), we consider the expectation of $\|\hbbeta_{\text{refitted}}-\bbeta_0\|_2^2$ by using the following decomposition:
\begin{align*}
E\|\hbbeta_{\text{refitted}} - \bbeta_0\|_2^2 & = E \{1_{\mathcal{E}} \|\hbbeta_{\text{refitted}} - \bbeta_0\|_2^2\} + E \{1_{\mathcal{E}^c} \|\hbbeta_{\text{refitted}} - \bbeta_0\|_2^2\}\\
&\leq E \{\bmu\t\bmu+2\bmu\t A\t \bveps+\bveps\t AA\t \bveps\} + E \{1_{\mathcal{E}^c} \|\hbbeta_{\text{refitted}}-\bbeta_0\|_2^2\}.
\end{align*}
Since $P(\mathcal{E}^c) = o(1)$ by Lemma \ref{L2} in Section \ref{SecSA.2} of Supplementary Material, the above inequality becomes an equation asymptotically by the dominated convergence theorem, which provides the basis for determining the orders of the risks. To ease the presentation, we do not distinguish between these two representations hereafter. The above decomposition, along with (\ref{017}), Condition \ref{cond1} and $\bveps \sim N(\textbf{0},\sigma^2 I_n)$, gives
\begin{align}\label{risk}
E\|\hbbeta_{\text{refitted}}-\bbeta_0\|_2^2 & \leq \bmu\t\bmu+\sigma^2 \tr( AA\t ) + E \{1_{\mathcal{E}^c} \|\hbbeta_{\text{refitted}}-\bbeta_0\|_2^2\} = I_1(\lambda_1) + I_2(\lambda_1) + I_3(\lambda_1),
\end{align}
where
\begin{align}
& I_1(\lambda_1) = \lambda_1^2 \bbeta_0\t(\bX_0\t \bX_0 + \lambda_1 I_s)^{-2} \bbeta_0, \\
& I_2(\lambda_1) = \sigma^2 \tr(\bX_0(\bX_0\t \bX_0 + \lambda_1 I_s)^{-2})\bX_0\t), \\
& I_3(\lambda_1) = E \{1_{\mathcal{E}^c} \|\hbbeta_{\text{refitted}} - \bbeta_0\|_2^2\}.
\end{align}

We analyze the first two terms, $I_1(\lambda_1)$ and $I_2(\lambda_1)$ in (\ref{risk}), by singular value decomposition. Since $\bX_0\t \bX_0$ is symmetric and positive semidefinite, there exists $s \times s$ orthonormal matrix $P$ such that $\bX_0\t \bX_0 = P\t D P$, where $D$ is a diagonal matrix with nonnegative elements $d_i, i = 1,\cdots, s$, the eigenvalues of $\bX_0\t \bX_0$. Replacing $\bX_0\t \bX_0$ with $P\t D P$, we get
\[\bX_0\t \bX_0 + \lambda_1 I_s = P\t (D + \lambda_1 I_s) P\ \text{ and }\ (\bX_0\t \bX_0 + \lambda_1 I_s)^{-2} = P\t (D + \lambda_1 I_s)^{-2} P.\]
Set $ \bb = (b_1,\cdots,b_s)\t=P \bbeta_0$. Then $\|\bb\|_2 = \|\bbeta_0\|_2$ and the first term becomes
\[I_1(\lambda_1) 
= \lambda_1^2 \bbeta_0\t P\t (D + \lambda_1 I_{s})^{-2} P \bbeta_0 = \sum_{i=1}^s \frac{\lambda_1^2 b_i^2}{(d_i+\lambda_1)^2} \]
and the second term can be simplified as
\begin{align*}
I_2(\lambda_1)& = \sigma^2 \tr(\bX_0\t \bX_0(\bX_0\t \bX_0 + \lambda_1 I_s)^{-2}) = \sigma^2 \tr(P\t D P P\t (D + \lambda_1 I_s)^{-2} P)\\
& = \sigma^2 \tr(D(D + \lambda_1 I_{s})^{-2}) = \sum_{i=1}^{s} \frac{\sigma^2 d_i}{(d_i + \lambda_1)^2}.
\end{align*}
Substituting the above two terms into (\ref{risk}), we get $E\|\hbbeta_{\text{refitted}}-\bbeta_0\|_2^2 \leq f(\lambda_1) + I_3(\lambda_1)$, where
\begin{equation}\label{flam}
f(\lambda_1)=\sum_{i=1}^s \frac{\lambda_1^2 b_i^2}{(d_i+\lambda_1)^2} + \sum_{i=1}^s \frac{\sigma^2 d_i}{(d_i+\lambda_1)^2}.
\end{equation}
Note that $f(\lambda_1)$ is a sum of two terms, with the first term increasing with $\lambda_1$ and the second term decreasing with $\lambda_1$. Besides $f(\lambda_1)$, we have another term $E \{1_{\mathcal{E}^c} \|\hbbeta_{\text{refitted}}-\bbeta_0\|_2^2\}$, which will be shown to be of a strictly smaller order than $f(\lambda_1)$.

\medskip

\noindent \textbf{Part 1.1: Identifying orders of optimal $\lambda_1$ and corresponding $f(\lambda_1)$ for $L_2$-risk}. It is hard to find the exact $\lambda_1$ minimizes $f(\lambda_1)$ since the denominators in the sum are different, but we can surely identify its order, in the following three steps.

First of all, we claim that $c^2 n \leq d_i \leq c_3^2 n$ for all $i$. It suffices to show that the maximum and minimum eigenvalues of $\bX_0\t \bX_0$, denoted as $\lambda_{\max}$ and $\lambda_{\min}$, can be bounded as $c^2 \leq \lambda_{\min}/n \leq \lambda_{\max}/n \leq c_3^2$. To this end, note that as $\bX_0$ is the submatrix of $\bX$ formed by columns with indices in $\supp(\bbeta_0)$ and $|\supp(\bbeta_0)| = s < M/2$ by Condition \ref{cond2}, we have $\lambda_{\min}/n \geq c^2$  by the property of robust spark. On the other hand, since we assumed $|\supp(\bbeta_0)| < M/2$, Condition \ref{cond4} ensures that $\lambda_{\max}/n \leq c_3^2$. So we have proved $c^2 \leq \lambda_{\min}/n \leq \lambda_{\max}/n \leq c_3^2$. Since $d_i$'s are the eigenvalues of $\bX_0\t \bX_0$, it follows that
\begin{align}\label{022}
c^2n \leq \lambda_{\min} \leq d_i \leq \lambda_{\max} \leq c_3^2n.
\end{align}
In fact, the same argument applies for any submatrix of $\bX$ with the number of columns less than $M/2$. Since $|\supp(\hbbeta_1)| \leq \|\hbbeta\|_0 < M/2$ by (\ref{002}),  we also have
\begin{equation}\label{023}
c^2 n \leq \lambda_{\min}(\bX_1\t \bX_1) \leq \lambda_{\max}(\bX_1\t \bX_1) \leq c_3^2 n,
\end{equation}
which will be used later for analyzing $I_3(\lambda_1)$.

Second, we show that the optimal $\lambda_1$ that minimizes $f(\lambda_1)$, denoted as $\lambda_{1,\text{opt}}$, is of the order $o(n)$. If it is not true, then there exists some constant $k > 0$ such that $\lambda_{1,\text{opt}} \geq kn$. By (\ref{022}), Condition \ref{cond4}, and since $\|\bbeta_0\|_2 \geq O(1)$, we have
\begin{equation}\label{on}
f(\lambda_{1,\text{opt}}) \geq \sum_{i=1}^s \frac{\lambda_{1,\text{opt}}^2 b_i^2}{(d_i+\lambda_{1,\text{opt}})^2} \geq \sum_{i=1}^s \frac{k^2 n^2 b_i^2}{(c_3^2 n + k n)^2} = \sum_{i=1}^s \frac{k^2 b_i^2}{(c_3^2 + k)^2} = \frac{k^2 \|\bbeta_0\|_2^2}{(c_3^2 + k)^2} \geq O(1).
\end{equation}
However, by the optimality of $\lambda_{1,\text{opt}}$, $f(\lambda_{1,\text{opt}}) \leq f(0) = \sum_{i=1}^s \frac{\sigma^2}{d_i} = O(\frac{s}{n}) = o(1)$. It is a contradiction and thus we must have $\lambda_{1,\text{opt}} = o(n)$.

In the third step, we go one step further to show that the order of $\lambda_{1,\text{opt}}$ is indeed $O(s \|\bbeta_0\|_2^{-2}) + O(s^2 n^{-1} \|\bbeta_0\|_2^{-4})$ 
by applying Taylor expansion on $f'(\lambda_1)$ with $\lambda_{1,\text{opt}} = o(n)$. Direct calculations yield
\begin{equation}\label{deriv}
f'(\lambda_1) = \sum_{i=1}^s \frac{2\lambda_1 b_i^2 d_i}{(d_i+\lambda_1)^3} - \sum_{i=1}^s \frac{2\sigma^2 d_i}{(d_i+\lambda_1)^3} = \sum_{i=1}^s \frac{2\lambda_1 b_i^2 d_i - 2\sigma^2 d_i}{(d_i+\lambda_1)^3}.
\end{equation}
Since the optimal $\lambda_1$ satisfies $f'(\lambda_{1,\text{opt}}) = 0$, we have
\begin{equation} \label{020}
\sum_{i=1}^s \frac{\lambda_{1,\text{opt}} b_i^2 d_i}{(d_i+\lambda_{1,\text{opt}})^3} = \sum_{i=1}^s \frac{\sigma^2 d_i}{(d_i+\lambda_{1,\text{opt}})^3}.
\end{equation}

We will rearrange the above equation as a quadratic equation for $\lambda_{1,\text{opt}}$ by using Taylor expansion. Since it has been proved that $\lambda_{1,\text{opt}} = o(n)$, or equivalently, $\lambda_{1,\text{opt}} = o(d_i)$ for each $1 \leq i \leq s$, we can apply Taylor expansion with Lagrange remainder to deal with the two fractions in (\ref{020}). For the left hand side of (\ref{020}), we have
\[\sum_{i=1}^s \frac{\lambda_{1,\text{opt}} b_i^2 d_i}{(d_i+\lambda_{1,\text{opt}})^3} = \sum_{i=1}^s \lambda_{1,\text{opt}} b_i^2 d_i (\frac{1}{d_i^3} - \frac{3 \lambda_{1,\text{opt}}}{(d_i + \omega_i)^4}) = \lambda_{1,\text{opt}}(\sum_{i=1}^s \frac{b_i^2}{d_i^2} - \sum_{i=1}^s \frac{3b_i^2 d_i \lambda_{1,\text{opt}}}{(d_i + \omega_i)^4}),\]
where $\omega_i$'s are numbers between $0$ and $\lambda_{1,\text{opt}}$. For the right hand side of (\ref{020}), we get
\[\sum_{i=1}^s \frac{\sigma^2 d_i}{(d_i+\lambda_{1,\text{opt}})^3} = \sum_{i=1}^s \sigma^2 d_i (\frac{1}{d_i^3} - \frac{3 \lambda_{1,\text{opt}}}{d_i^4} + \frac{6 \lambda_{1,\text{opt}}^2}{(d_i + \gamma_i)^5}) = \sum_{i=1}^s \frac{\sigma^2}{d_i^2} - \lambda_{1,\text{opt}} \sum_{i=1}^s \frac{3\sigma^2}{d_i^3} + \sum_{i=1}^s \frac{6 \sigma^2 d_i \lambda_{1,\text{opt}}^2}{(d_i + \gamma_i)^5},\]
where $\gamma_i$'s are numbers between $0$ and $\lambda_{1,\text{opt}}$. Equalling the two sides yields
\[\lambda_{1,\text{opt}}(\sum_{i=1}^s \frac{b_i^2}{d_i^2} - \sum_{i=1}^s \frac{3b_i^2 d_i \lambda_{1,\text{opt}}}{(d_i + \omega_i)^4}) = \sum_{i=1}^s \frac{\sigma^2}{d_i^2} - \lambda_{1,\text{opt}} \sum_{i=1}^s \frac{3\sigma^2}{d_i^3} + \sum_{i=1}^s \frac{6 \sigma^2 d_i \lambda_{1,\text{opt}}^2}{(d_i + \gamma_i)^5}.\]
Reorganizing it in terms of the power of $\lambda_{1,\text{opt}}$, we obtain:
\begin{equation}\label{equa}
\left(\sum_{i=1}^s \frac{6 \sigma^2 d_i}{(d_i + \gamma_i)^5} + \sum_{i=1}^s \frac{3b_i^2 d_i}{(d_i + \omega_i)^4} \right) \lambda_{1,\text{opt}}^2 - \left(\sum_{i=1}^s \frac{b_i^2}{d_i^2} + \sum_{i=1}^s \frac{3\sigma^2}{d_i^3}\right) \lambda_{1,\text{opt}} + \sum_{i=1}^s \frac{\sigma^2}{d_i^2} = 0.
\end{equation}
Its solution for $\lambda_{1,\text{opt}}$ is $\frac{-b - \sqrt{b^2 - 4ac}}{2a}$, where $a = \sum_{i=1}^s \frac{6 \sigma^2 d_i}{(d_i + \gamma_i)^5} + \sum_{i=1}^s \frac{3b_i^2 d_i}{(d_i + \omega_i)^4}, b = -(\sum_{i=1}^s \frac{b_i^2}{d_i^2} + \sum_{i=1}^s \frac{3\sigma^2}{d_i^3})$ and $c = \sum_{i=1}^s \frac{\sigma^2}{d_i^2}$. We drop the solution $\lambda_{1,\text{opt}} = \frac{-b + \sqrt{b^2 - 4ac}}{2a}$ since its order is $O(n)$, which can be proved by analyzing the orders of $a,b$ and $c$ as follows.

With $c^2 n \leq d_i \leq c_3^2 n$, we can immediately calculate the orders of terms in $a,b$ and $c$ as
\begin{align*}
\sum_{i=1}^s \frac{6 \sigma^2 d_i}{(d_i + \gamma_i)^5} = & \ O(sn^{-4}), \ \ \sum_{i=1}^s \frac{3b_i^2 d_i}{(d_i + \omega_i)^4} = O(n^{-3} \|\bbeta_0\|_2^2), \ \ \sum_{i=1}^s \frac{b_i^2}{d_i^2} = O(n^{-2} \|\bbeta_0\|_2^2),\\
&\sum_{i=1}^s \frac{3\sigma^2}{d_i^3} = O(sn^{-3}), \ \ \sum_{i=1}^s \frac{\sigma^2}{d_i^2} = O(sn^{-2}).
\end{align*}
Then we have $a = O(n^{-3} \|\bbeta_0\|_2^2)$, $b = O(n^{-2} \|\bbeta_0\|_2^2)$, and $c = O(sn^{-2})$. We know that $b^2 = O(n^{-4} \|\bbeta_0\|_2^4)$ is the leading term in $b^2 - 4ac$ since $4ac = O(s n^{-5} \|\bbeta_0\|_2^2)$. Since $b < 0$, both $- b$ and $\sqrt{b^2 - 4ac}$ are positive and they are of the same order $O(n^{-2} \|\bbeta_0\|_2^2)$. So the order for $\frac{-b + \sqrt{b^2 - 4ac}}{2a}$ is $O(n^{-2} \|\bbeta_0\|_2^2)/O(n^{-3} \|\bbeta_0\|_2^2) = O(n)$. Since we have proved $\lambda_{1,\text{opt}} = o(n)$ before, this rules out the possibility of $\lambda_{1,\text{opt}} = \frac{-b + \sqrt{b^2 - 4ac}}{2a}$, which entails that $\lambda_{1,\text{opt}} = \frac{-b - \sqrt{b^2 - 4ac}}{2a}$.
We further show that $\lambda_{1,\text{opt}}$ has a leading order $O(s \|\bbeta_0\|_2^{-2})$ followed by a secondary order $O(s^2 n^{-1} \|\bbeta_0\|_2^{-4})$, in Section \ref{SecSB.1} of Supplementary Material.

Plugging $\lambda_{1,\text{opt}} = O(s \|\bbeta_0\|_2^{-2}) + O(s^2 n^{-1} \|\bbeta_0\|_2^{-4})$ into $f(\lambda_1)$ defined in (\ref{flam}), we obtain
\[\sum_{i=1}^s \frac{\lambda_{1,\text{opt}}^2 b_i^2}{(d_i+\lambda_{1,\text{opt}})^2} = O(\frac{s^2}{n^2 \|\bbeta_0\|_2^2}) + O(\frac{s^3}{n^3 \|\bbeta_0\|_2^4}),\ \ \sum_{i=1}^s \frac{\sigma^2 d_i}{(d_i+\lambda_{1,\text{opt}})^2} = O(\frac{s}{n}) + O(\frac{s^2}{n^2 \|\bbeta_0\|_2^2}).\]
Thus, the order for $f(\lambda_{1,\text{opt}})$ is $O(s/n) + O(s^2 n^{-2} \|\bbeta_0\|_2^{-2})$.

\medskip

\noindent \textbf{Part 1.2: Bounding the leading term of order $O(s \|\bbeta_0\|_2^{-2})$ in $\lambda_{1,\emph{opt}}$}. In fact, the leading order $O(s \|\bbeta_0\|_2^{-2})$ in $\lambda_{1,\text{opt}}$ comes from $- t/(2a)$, which equals to $-c/b$ since $4ac = 2bt$. Plugging the definitions of $b$ and $c$ gives
\[
-\frac{c}{b} = \frac{\sum_{i=1}^s \frac{\sigma^2}{d_i^2}}{\sum_{i=1}^s \frac{b_i^2}{d_i^2} + \sum_{i=1}^s \frac{3\sigma^2}{d_i^3}}.
\]
By (\ref{022}), we see that $\sum_{i=1}^s \frac{3\sigma^2}{d_i^3}$ is a smaller order term compared with $\sum_{i=1}^s \frac{b_i^2}{d_i^2}$. Thus, the leading term for  $-\frac{c}{b}$ is $(\sum_{i=1}^s\frac{\sigma^2}{d_i^2})(\sum_{i=1}^s \frac{b_i^2}{d_i^2})^{-1}$.

Recall that $\lambda_{\min}$ and $\lambda_{\max}$ stand for the smallest and largest eigenvalues of $\bX_0\t \bX_0$. With $\lambda_{\min} \leq d_i \leq \lambda_{\max}$ and $\sum_{i=1}^s b_i^2 = \|\bbeta_0\|_2^2$, we obtain that the leading term for $-c/b$ can be bounded as
\[\frac{s\sigma^2}{\|\bbeta_0\|_2^2} \frac{\lambda_{\min}^2}{\lambda_{\max}^2}=\frac{\sum_{i=1}^s \frac{\sigma^2}{\lambda_{\max}^2}}{\sum_{i=1}^s \frac{b_i^2}{\lambda_{\min}^2}} \leq \frac{\sum_{i=1}^s \frac{\sigma^2}{d_i^2}}{\sum_{i=1}^s \frac{b_i^2}{d_i^2}} \leq \frac{\sum_{i=1}^s \frac{\sigma^2}{\lambda_{\min}^2}}{\sum_{i=1}^s \frac{b_i^2}{\lambda_{\max}^2}}= \frac{s\sigma^2}{\|\bbeta_0\|_2^2} \frac{\lambda_{\max}^2}{\lambda_{\min}^2}.\]

So the leading term for $\lambda_{1,\text{opt}}$, which is $O(s \|\bbeta_0\|_2^{-2})$, is between $\frac{s\sigma^2}{\|\bbeta_0\|_2^2} \frac{\lambda_{\min}^2}{\lambda_{\max}^2}$ and $\frac{s\sigma^2}{\|\bbeta_0\|_2^2} \frac{\lambda_{\max}^2}{\lambda_{\min}^2}$. In particular, when $\lambda_{\max}$ equals to $\lambda_{\min}$, which implies all $d_i$'s are the same, we can solve (\ref{deriv}) readily to get $\lambda_{1,\text{opt}} = \frac{s\sigma^2}{\|\bbeta_0\|_2^2}$, which coincides with the above bounds for the leading term.

\medskip

\noindent \textbf{Part 1.3: Bounding term $E \{1_{\mathcal{E}^c} \|\hbbeta_{\text{refitted}}-\bbeta_0\|_2^2\}$ in (\ref{risk})}. Now let us turn to the last term in (\ref{risk}): $I_3(\lambda_1) = E \{1_{\mathcal{E}^c} \|\hbbeta_{\text{refitted}}-\bbeta_0\|_2^2\}$. We 
prove in Section \ref{SecSB.2} of Supplementary Material that compared with $f(\lambda_{1,\text{opt}})$, the order of $E \{1_{\mathcal{E}^c} \|\hbbeta_{\text{refitted}}-\bbeta_0\|_2^2\}$ is much smaller. 
Thus $f(\lambda_{1,\text{opt}})$ is the leading term of 
$E\|\hbbeta_{\text{refitted}}-\bbeta_0\|_2^2$ and
\begin{equation}\label{028}
E\|\hbbeta_{\text{refitted}}-\bbeta_0\|_2^2 = O(s/n) + O(s^2 n^{-2} \|\bbeta_0\|_2^{-2})
\end{equation}
for the optimal choice of $\lambda_1$.

\medskip

\noindent \textbf{Part 1.4: Bounds for the $L_q$-risks}. Based on the risk for squared $L_2$-loss above, we can derive the bounds for the risks of $L_q$-losses by using H\"{o}lder's inequality, as shown in Section \ref{SecSB.3} of Supplementary Material.
The bound under $L_{\infty}$-loss follows directly from the inequality
$E\|\hbbeta_{\text{refitted}}-\bbeta_0\|_{\infty} \leq E\|\hbbeta_{\text{refitted}}-\bbeta_0\|_2$.

\medskip

\noindent \textbf{Part 2: Risk properties for $\hbbeta_{\text{refitted}}$ under prediction loss}. In this part, we will find the risk property of the prediction loss for the refitted estimator $\hbbeta_{\text{refitted}}$ in a very similar way as before.

Similarly to (\ref{017}), we have $\|\bX(\hbbeta_{\text{refitted}}-\bbeta_0)\|_2^2  = (\bmu\t+ \bveps\t A)\bX_0\t \bX_0(\bmu+A\t \bveps) = \bmu\t \bX_0\t \bX_0 \bmu + 2\bmu\t \bX_0\t \bX_0 A\t \bveps + \bveps\t A \bX_0\t \bX_0 A\t \bveps$. Taking expectation to the prediction loss, we have
\begin{align} \label{expect}
&n^{-1}E\|\bX(\hbbeta_{\text{refitted}}-\bbeta_0)\|_2^2=  n^{-1}E \{1_{\mathcal{E}} \|\bX(\hbbeta_{\text{refitted}}-\bbeta_0)\|_2^2\} + n^{-1}E \{1_{\mathcal{E}^c} \|\bX(\hbbeta_{\text{refitted}}-\bbeta_0)\|_2^2\} \nonumber\\
\leq \ & n^{-1}E \left\{(\bmu\t \bX_0\t \bX_0 \bmu + 2\bmu\t \bX_0\t \bX_0 A\t \bveps + \bveps\t A \bX_0\t \bX_0 A\t \bveps)\right\} + n^{-1}E \left\{1_{\mathcal{E}^c} \|\bX(\hbbeta_{\text{refitted}}-\bbeta_0)\|_2^2\right\} \nonumber\\
= \ & n^{-1}(\bmu\t \bX_0\t \bX_0 \bmu + \sigma^2 \tr(A \bX_0\t \bX_0 A\t)) + n^{-1}E \{1_{\mathcal{E}^c} \|\bX(\hbbeta_{\text{refitted}}-\bbeta_0)\|_2^2\}.
\end{align}

Using definitions $\bmu=-\lambda_1(\bX_0\t \bX_0 + \lambda_1 I_s)^{-1} \bbeta_0$, $A=\bX_0 (\bX_0\t \bX_0 + \lambda_1 I_s)^{-1}$ and $\bX_0\t \bX_0 = P\t D P$, we get
\begin{align*}
\bmu\t \bX_0\t \bX_0 \bmu = \lambda_1^2 (P \bbeta_0)\t (D + \lambda_1 I_s)^{-1} D (D + \lambda_1 I_s)^{-1} P \bbeta_0 = \lambda_1^2 \sum_{i=1}^s \frac{b_i^2 d_i}{(d_i + \lambda_1)^2},
\end{align*}
and
\begin{align*}
\sigma^2 \tr(A \bX_0\t \bX_0 A\t)& = \sigma^2 \tr(P\t D P P\t (D + \lambda_1 I_s)^{-1} P P\t D P P\t (D + \lambda_1 I_s)^{-1} P)\\
&= \sigma^2 \tr(D(D + \lambda_1 I_s)^{-1}D(D + \lambda_1 I_s)^{-1})= \sum_{i=1}^s \frac{\sigma^2 d_i^2}{(d_i+\lambda_1)^2}.
\end{align*}
Plugging the above two terms into (\ref{expect}) yields
\begin{equation*}
\frac{1}{n} E\|\bX(\hbbeta_{\text{refitted}}-\bbeta_0)\|_2^2 \leq \frac{1}{n}\left(\lambda_1^2 \sum_{i=1}^s \frac{b_i^2 d_i}{(d_i + \lambda_1)^2} + \sum_{i=1}^s \frac{\sigma^2 d_i^2}{(d_i+\lambda_1)^2}\right) + \frac{1}{n}E \left\{1_{\mathcal{E}^c} \|\bX(\hbbeta_{\text{refitted}}-\bbeta_0)\|_2^2\right\}.
\end{equation*}

Set
\[g(\lambda_1) = \lambda_1^2 \sum_{i=1}^s \frac{b_i^2 d_i}{(d_i + \lambda_1)^2} + \sum_{i=1}^s \frac{\sigma^2 d_i^2}{(d_i+\lambda_1)^2},\]
and note that it can be transformed from $f(\lambda_1)$ by multiplying $d_i$ in the $i$th term of each sum. Denote the optimal $\lambda_1$ for minimizing $g(\lambda_1)$ as $\lambda'_{1,\text{opt}}$. In view of (\ref{on}) and (\ref{equa}), the same argument applies, we also get $\lambda'_{1,\text{opt}} = o(n)$ and consequently, we can deduce $\lambda'_{1,\text{opt}} = O(s \|\bbeta_0\|_2^{-2}) + O(s^2 n^{-1} \|\bbeta_0\|_2^{-4})$ as the ratio of orders does not change. Then we can prove that the leading term for $\lambda'_{1,\text{opt}}$ is between $\frac{s\sigma^2}{\|\bbeta_0\|_2^2} \frac{\lambda_{\min}}{\lambda_{\max}}$ and $\frac{s\sigma^2}{\|\bbeta_0\|_2^2} \frac{\lambda_{\max}}{\lambda_{\min}}$ and $g(\lambda'_{1,\text{opt}}) = O(s) + O(s^2 n^{-1} \|\bbeta_0\|_2^{-2})$. The term $n^{-1}E \{1_{\mathcal{E}^c} \|\bX(\hbbeta_{\text{refitted}}-\bbeta_0)\|_2^2\}$ can be shown to have a smaller order than $O(s^2 n^{-2} \|\bbeta_0\|_2^{-2})$ similarly as before. Therefore, $n^{-1}E\|\bX(\hbbeta_{\text{refitted}}-\bbeta_0)\|_2^2 = O(s/n) + O(s^2 n^{-2} \|\bbeta_0\|_2^{-2})$, which concludes the proof.

\newpage

\setcounter{page}{1}
\setcounter{section}{0}
\setcounter{equation}{0}

\renewcommand{\theequation}{A.\arabic{equation}}
\setcounter{equation}{0}

\begin{center}{\bf \large Supplementary Material to ``High-Dimensional Thresholded Regression and Shrinkage Effect''}

\bigskip

Zemin Zheng, Yingying Fan and Jinchi Lv
\end{center}

\noindent This Supplementary Material contains proofs of Lemmas \ref{L1}--\ref{L2}, and technical details in the proof of Theorem \ref{Thm5}.

\section{Proofs of Lemmas} \label{SecSA}

\subsection{Proof of Lemma \ref{L1}} \label{SecSA.1}
Since each covariate vector $\bx_j$ is rescaled to have $L_2$-norm $n^{1/2}$, the solution $\hbeta(z) = z 1_{\{|z| > \lambda\}}$ with $z = n^{-1} (\by - \bX \bbeta_j)\t  \bx_j$ can be easily derived for the univariate penalized least-squares estimator for both penalties $p_{H, \lambda}(t)$ and $p_{H_0, \lambda}(t)$. 

\subsection{Lemma \ref{L2} and its proof} \label{SecSA.2}
\begin{lemma} \label{L2}
Define two events
\begin{align}
\mathcal{E} & = \{\|n^{-1} \bX\t \bveps\|_{\infty} \leq c_2 \sqrt{(\log \widetilde{p})/n} \} \ \text{ and } \ \mathcal{E'}  = \{\|n^{-1} \bX_0\t \bveps\|_{\infty} \leq c'_2 \sqrt{(\log n)/n}\} \label{event}
\end{align}
with $c_2 \geq \sqrt{10}\sigma$ and $c_2' \geq \sqrt{2}\sigma$ some positive constants. Then we have
\begin{align*}
& P(\mathcal{E}^c)\leq  (2/\pi)^{1/2}c_2^{-1}\sigma(\log \widetilde{p})^{-1/2}\widetilde{p}^{1-\frac{c_2^2}{2\sigma^2}}\rightarrow 0, \\
& P(\mathcal{E'}^c)\leq  (2/\pi)^{1/2} c_2'^{-1} \sigma s (\log n)^{-1/2} n^{-\frac{c_2'^2}{2 \sigma^2}} \rightarrow 0,
\end{align*}
as $n \rightarrow \infty$.
\end{lemma}

\noindent \textit{Proof of Lemma \ref{L2}}: The proofs for the inequalities on  $P(\mathcal{E}^c)$  and $P(\mathcal{E'}^c)$ are similar, so we only outline the first one. Since the $j$-th covariate vector $\bx_j$ has been rescaled to have $L_2$-norm $n^{1/2}$ and $\bveps \sim N(\textbf{0},\sigma^2 I_n)$, we have $n^{-1} \bx_j\t \bveps \sim N(0, \sigma^2/n)$ for each $j$. By Bonferroni's inequality and Gaussian tail probability bound (see Proposition 2.2.1 in Dudley, 1999), we have
\begin{align*}
P(\mathcal{E}^c) & \leq \sum^{p}_{j=1} P\Big (|n^{-1} \bx_j\t \bveps| > c_2 \sqrt{(\log \widetilde{p})/n} \Big )\leq \sum^{p}_{j=1} \frac{2\sigma}{c_2 \sqrt{\log \widetilde{p}}} \frac{1}{\sqrt{2 \pi}} e^{-\frac{c_2^2 \log \widetilde{p}}{2\sigma^2}} = \frac{\sqrt{2}\sigma}{c_2 \sqrt{\pi \log \widetilde{p}}} \widetilde{p}^{1-\frac{c_2^2}{2\sigma^2}},
\end{align*}
which tends to $0$ as $n\to \infty$ since $c_2 \geq \sqrt{10}\sigma$.

\section{Technical details in the proof of Theorem \ref{Thm5}} \label{SecSB}

\subsection{Identifying the order of $\lambda_{1,\text{opt}}$} \label{SecSB.1}
We find the order of $\lambda_{1,\text{opt}}$ by analyzing $\frac{-b - \sqrt{b^2 - 4ac}}{2a}$. To this end, we
first find out the order of $t$ that satisfies $4ac = 2bt$. By direct calculations, we have
\begin{align*}
4ac & = 4\left\{\sum_{i=1}^s \frac{6 \sigma^2 d_i}{(d_i + \gamma_i)^5} + \sum_{i=1}^s \frac{3b_i^2 d_i}{(d_i + \omega_i)^4}\right\}(\sum_{i=1}^s \frac{\sigma^2}{d_i^2}) \\
& = 4\left\{\sum_{i=1}^s \frac{6s \sigma^2 d_i}{(d_i + \gamma_i)^5 \|\bbeta_0\|_2^2} + \sum_{i=1}^s \frac{3s b_i^2 d_i}{(d_i + \omega_i)^4 \|\bbeta_0\|_2^2}\right\}(\sum_{i=1}^s \frac{\sigma^2 \|\bbeta_0\|_2^2}{s d_i^2}),
\end{align*}
where $\sum_{i=1}^s \frac{\sigma^2 \|\bbeta_0\|_2^2}{s d_i^2}$ is of the same order as $b$, and thus, $t$ should have order
\[4\left\{\sum_{i=1}^s \frac{6s \sigma^2 d_i}{(d_i + \gamma_i)^5 \|\bbeta_0\|_2^2} + \sum_{i=1}^s \frac{3s b_i^2 d_i}{(d_i + \omega_i)^4 \|\bbeta_0\|_2^2}\right\} = O(s^2 n^{-4} \|\bbeta_0\|_2^{-2}) + O(sn^{-3}).\]

Replacing $4ac$ with $2bt$ and by Taylor expansion, we have
\[\sqrt{b^2 - 4ac} = \sqrt{b^2 - 2bt} = \sqrt{(b - t)^2 - t^2} = -(b - t) - \frac{t^2}{2\sqrt{(b - t)^2 - t'}},\]
where $t'$ is a number between $0$ and $t^2$. By the above calculations on the order of $t$, we have
\[
\frac{t^2}{2\sqrt{(b - t)^2 - t'}} = O(s^2 n^{-6})/O(n^{-2} \|\bbeta_0\|_2^2) = O(s^2 n^{-4} \|\bbeta_0\|_2^{-2}).
\]
Thus, combining the above two results, the order of optimal $\lambda$ can be calculated as follows:
\begin{align*}
\lambda_{1,\text{opt}} &= \frac{-b - \sqrt{b^2 - 4ac}}{2a} = \frac{1}{2a}(- t + \frac{t^2}{2\sqrt{(b - t)^2 - t'}})\\
& = O(n^3 \|\bbeta_0\|_2^{-2}) (O(sn^{-3}) + O(s^2 n^{-4} \|\bbeta_0\|_2^{-2}) + O(s^2 n^{-4} \|\bbeta_0\|_2^{-2}))\\
&= O(s \|\bbeta_0\|_2^{-2}) + O(s^2 n^{-1} \|\bbeta_0\|_2^{-4}).
\end{align*}
Therefore, $\lambda_{1,\text{opt}}$ has a leading order $O(s \|\bbeta_0\|_2^{-2})$ followed by a secondary order $O(s^2 n^{-1} \|\bbeta_0\|_2^{-4})$.

\subsection{Bounding term $I_3(\lambda_1)$} \label{SecSB.2}
We 
proceed to bound $I_3(\lambda_1) = E \{1_{\mathcal{E}^c} \|\hbbeta_{\text{refitted}}-\bbeta_0\|_2^2\}$. Since $\|\hbbeta_{\text{refitted}}-\bbeta_0\|_2^2 \leq 2(\|\hbbeta_{\text{refitted}}\|_2^2 + \|\bbeta_0\|_2^2)$, we have
\begin{equation}\label{026}
E \{1_{\mathcal{E}^c} \|\hbbeta_{\text{refitted}}-\bbeta_0\|_2^2\} \leq E \{1_{\mathcal{E}^c} 2(\|\hbbeta_{\text{refitted}}\|_2^2 + \|\bbeta_0\|_2^2)\} = 2E (1_{\mathcal{E}^c} \|\hbbeta_{\text{refitted}}\|_2^2) + 2 P(\mathcal{E}^c) \|\bbeta_0\|_2^2.
\end{equation}
Note that $E (1_{\mathcal{E}^c} \|\bbeta_0\|_2^2) = O(P(\mathcal{E}^c) \|\bbeta_0\|_2^2) = O(\frac{1}{\sqrt{\log \widetilde{p}}}\widetilde{p}^{1-\frac{c_2^2}{2\sigma^2}} \|\bbeta_0\|_2^2) = o(\widetilde{p}^{1-\frac{c_2^2}{2\sigma^2}} \|\bbeta_0\|_2^2)$, which is much smaller than $O(s^2 n^{-2} \|\bbeta_0\|_2^{-2})$ since $c_2$ can be chosen arbitrarily large for any given level of signal strength $\|\bbeta_0\|_2$. So it remains to bound $E (1_{\mathcal{E}^c} \|\hbbeta_{\text{refitted}}\|_2^2)$.

We first bound $\|\hbbeta_{\text{refitted}}\|_2$ as
\[\|\hbbeta_{\text{refitted}}\|_2 \leq \left\|(\bX_1\t \bX_1 + \lambda_1 I_{s_1})^{-1} \bX_1\t \|_2 \|\by \right\|_2 = \left\{\lambda_{\max}(\bX_1(\bX_1\t \bX_1 + \lambda_1 I_{s_1})^{-2}\bX_1\t)\right\}^{1/2} \|\by\|_2.\]
By (\ref{023}), we have $\bX_1\t \bX_1 + \lambda_1 I_{s_1} \geq (c^2 n + \lambda_1)I_{s_1}$, where $\geq$ means $\bX_1\t \bX_1 + \lambda_1 I_{s_1} - (c^2 n + \lambda_1)I_{s_1}$ is positive semidefinite. It follows that
\[\lambda_{\max}(\bX_1(\bX_1\t \bX_1 + \lambda_1 I_{s_1})^{-2}\bX_1\t) \leq \frac{\lambda_{\max}(\bX_1\bX_1\t)}{(c^2 n + \lambda_1)^2} = \frac{O(n)}{(c^2 n + \lambda_1)^2}.\]
Then we have
\begin{equation}\label{024}
\|\hbbeta_{\text{refitted}}\|_2^2 \leq \lambda_{\max}(\bX_1(\bX_1\t \bX_1 + \lambda_1 I_{s_1})^{-2}\bX_1\t) \|\by\|_2^2 \leq \frac{O(n) \|\by\|_2^2}{(c^2 n + \lambda_1)^2},
\end{equation}
On the other hand, since $\|\by\|_2^4 = \|\bX_0 \bbeta_0 + \bveps\|_2^4 = (\bbeta_0\t \bX_0\t \bX_0 \bbeta_0 + 2\bbeta_0\t \bX_0\t \bveps + \bveps\t \bveps)^2$ and $\bveps \sim N(\textbf{0},\sigma^2 I_n)$, by (\ref{022}), we have
\begin{align}\label{025}
E\|\by\|_2^4 &= (\bbeta_0\t \bX_0\t \bX_0 \bbeta_0)^2 + 2(\bbeta_0\t \bX_0\t \bX_0 \bbeta_0)E(\bveps\t \bveps) + 4E(\bbeta_0\t \bX_0\t \bveps)^2 + E(\bveps\t \bveps)^2 \nonumber\\
&= O(n^2 \|\bbeta_0\|_2^4) + O(n^2 \|\bbeta_0\|_2^2) + O(n \|\bbeta_0\|_2^2) + O(n^2) = O(n^2 \|\bbeta_0\|_2^4)
\end{align}
where all of the four terms above are positive. Combining (\ref{024}) with (\ref{025})  and by the Cauchy-Schwarz inequality, we have
\begin{align*}
E\{1_{\mathcal{E}^c} \|\hbbeta_{\text{refitted}}\|_2^2\} &\leq \frac{O(n)}{(c^2 n + \lambda_1)^2} E \{1_{\mathcal{E}^c}\|\by\|_2^2\}\leq
\frac{O(n)}{(c^2 n + \lambda_1)^2}(E \{1_{\mathcal{E}^c}^2\})^{\frac{1}{2}} (E \|\by\|_2^4)^{\frac{1}{2}}
\\
&= O(P(\mathcal{E}^c)^{\frac{1}{2}} \|\bbeta_0\|_2^2)  = O\left\{(\log \widetilde{p})^{-\frac{1}{4}}\widetilde{p}^{\frac{1}{2}-\frac{c_2^2}{4\sigma^2}} \|\bbeta_0\|_2^2\right\} = o\left\{ \widetilde{p}^{\frac{1}{2}-\frac{c_2^2}{4\sigma^2}} \|\bbeta_0\|_2^2 \right\},
\end{align*}
which is also strictly smaller than $O(s^2 n^{-2} \|\bbeta_0\|_2^{-2})$ since $c_2$ can be chosen arbitrarily large for any given level of signal strength $\|\bbeta_0\|_2$. The above inequality together with (\ref{026}) and Lemma \ref{L2} ensures that
\begin{align} \label{boundE}
\nonumber E \{1_{\mathcal{E}^c} \|\hbbeta_{\text{refitted}}-\bbeta_0\|_2^2\} &\leq 2E (1_{\mathcal{E}^c} \|\hbbeta_{\text{refitted}}\|_2^2) + 2P(\mathcal{E}^c) \|\bbeta_0\|_2^2 \\
& \leq o(\widetilde{p}^{\frac{1}{2}-\frac{c_2^2}{4\sigma^2}} \|\bbeta_0\|_2^2) = o\left\{s^2 n^{-2} \|\bbeta_0\|_2^{-2}\right\}.
\end{align}

\subsection{Bounds for the $L_q$-risks with $q \in [1,2)$} \label{SecSB.3}
We first show that for $q \in [1,2)$, $E(1_{\mathcal{E}^c} \|\hbbeta_{\text{refitted}}-\bbeta_0\|_q^q)$ has a smaller order than $O(s^2 \|\bbeta_0\|_2^{-2}/n^{q/2 + 1})$ when we choose large enough $c_2$. 
Since by (\ref{002}) and $s < M/2$ in Condition \ref{cond2}, $\|\hbbeta_{\text{refitted}}-\bbeta_0\|_0 \leq \|\hbbeta_{\text{refitted}}\|_0 + \|\bbeta_0\|_0 \leq \frac{M}{2} + \frac{M}{2} = M \leq n + 1$, then H\"{o}lder's inequality ensures $\|\hbbeta_{\text{refitted}}-\bbeta_0\|_q^q \leq (n + 1)^{1-q/2} \|\hbbeta_{\text{refitted}}-\bbeta_0\|_2^q$. Thus, by Cauchy-Schwarz inequality and (\ref{boundE}) we have
\begin{align}\label{027}
\nonumber E(1_{\mathcal{E}^c} \|\hbbeta_{\text{refitted}}-\bbeta_0\|_q^q) & \leq (n + 1)^{1-q/2} E(1_{\mathcal{E}^c} \|\hbbeta_{\text{refitted}}-\bbeta_0\|_2^q) \leq (n + 1)^{1-q/2} E(1_{\mathcal{E}^c} \|\hbbeta_{\text{refitted}}-\bbeta_0\|_2^2)^{q/2} \\
& \leq 
o(n^{1-\frac{q}{2}} p^{\frac{q}{4}-\frac{c_2^2 q}{8\sigma^2}} \|\bbeta_0\|_2^q) \leq O(s^2 \|\bbeta_0\|_2^{-2}/n^{q/2 + 1}),
\end{align}
where the last step is because $c_2$ can be chosen sufficiently large.

By H\"{o}lder's inequality, when $q \in [1,2)$, we have $\|\hbbeta_{\text{refitted}}-\bbeta_0\|_q \leq s^{1/q-1/2} \|\hbbeta_{\text{refitted}}-\bbeta_0\|_2$ on event $\mathcal{E}$, which together with (\ref{027}) and the $L_2$-loss bound (\ref{028}) gives
\begin{align*}
& E\|\hbbeta_{\text{refitted}}-\bbeta_0\|_q^q\leq s^{1-q/2} E\|\hbbeta_{\text{refitted}}-\bbeta_0\|_2^q + E(1_{\mathcal{E}^c} \|\hbbeta_{\text{refitted}}-\bbeta_0\|_q^q) \\
& \leq s^{1-q/2}(E\|\hbbeta_{\text{refitted}}-\bbeta_0\|_2^2)^{q/2} + E(1_{\mathcal{E}^c} \|\hbbeta_{\text{refitted}}-\bbeta_0\|_q^q) =  O(s/n^{q/2}) + O(s^2 \|\bbeta_0\|_2^{-2}/n^{q/2 + 1}).
\end{align*}

\end{document}